\documentclass[fleqn,usenatbib]{mnras}

\usepackage{newtxtext,newtxmath}

\usepackage[T1]{fontenc}

\DeclareRobustCommand{\VAN}[3]{#2}
\let\VANthebibliography\thebibliography
\def\thebibliography{\DeclareRobustCommand{\VAN}[3]{##3}\VANthebibliography}

\newcommand{\cms}{\,$\text{cm}\,\text{s}^{-1}\,$}	
\newcommand{\ms}{\,$\text{m}\,\text{s}^{-1}\,$}	
\newcommand{\kms}{\,$\text{km}\,\text{s}^{-1}$}	
\newcommand{\bobs}{\,$|\hat{B}_{\text{obs}}|$\,}
\newcommand{\rhk}{\,$\log R'_{HK}$\,}
\newcommand{\dabb}{\,$\Delta\alpha B^2$\,}
\newcommand{\abb}{\,$\alpha B^2$\,}

\usepackage{graphicx}	
\usepackage{amsmath}	

\title[Unsigned magnetic flux proxy]{Unsigned magnetic flux proxy from solar optical intensity spectra}

\author[Lienhard et al.]{
F. Lienhard,$^{1}$\thanks{E-mail: fl386@cam.ac.uk}
A. Mortier,$^{1,2,3}$
H. M. Cegla,$^{4}$
A. Collier Cameron,$^{5}$
B. Klein,$^{6}$
C. A. Watson$^{7}$
\\
$^{1}$Astrophysics Group, Cavendish Laboratory, University of Cambridge, J.J. Thomson Avenue, Cambridge CB3 0HE, UK\\
$^{2}$Kavli Institute for Cosmology, University of Cambridge, Madingley Road, Cambridge CB3 0HA, UK\\
$^{3}$School of Physics \& Astronomy, University of Birmingham, Edgbaston, Birmingham B15 2TT, UK\\
$^{4}$ Department of Physics, University of Warwick, Gibbet Hill Road, Coventry CV4 7AL, UK\\
$^{5}$ SUPA School of Physics and Astronomy, University of St Andrews, North Haugh, St Andrews KY16 9SS, UK\\
$^{6}$ Department of Physics, University of Oxford, OX13RH, Oxford, UK\\
$^{7}$ Astrophysics Research Centre, School of Mathematics and Physics, Queen's University Belfast, BT7 1NN, Belfast, UK
}

\date{Accepted 2023 April 28. Received 2023 April 28; in original form 2022 October 21}

\pubyear{2023}

\begin{document}
\label{firstpage}
\pagerange{\pageref{firstpage}--\pageref{lastpage}}
\maketitle

\begin{abstract}
The photospheric unsigned magnetic flux has been shown to be highly correlated with radial velocity (RV) variations caused by solar surface activity. This activity indicator is therefore a prime candidate to unlock the potential of RV surveys to discover Earth twins orbiting Sun-like stars.
We show for the first time how a precise proxy of the unsigned magnetic flux (\dabb) can be obtained from Sun-as-a-star intensity spectra by harnessing the magnetic information contained in over 4000 absorption lines in the wavelength range from 380 to 690 nm. This novel activity proxy can thus be obtained from the same spectra from which RVs are routinely extracted.
We derived \dabb from 500 randomly selected spectra from the HARPS-N public solar data set, which spans from 2015 to 2018. We compared our estimates with the unsigned magnetic flux values from the Solar Dynamics Observatory (SDO) finding excellent agreement (median absolute deviation: 4.9 per cent).
The extracted indicator \dabb correlates with SDO's unsigned magnetic flux estimates on the solar rotational timescale (Pearson correlation coefficient 0.67) and on the three-year timescale of our data set (correlation coefficient 0.91). We find correlations of \dabb with the HARPS-N solar RV variations of 0.49 on the rotational timescale and 0.78 on the three-year timescale.
The Pearson correlation of \dabb with the RVs is found to be greater than the correlation of the classical activity indicators with the RVs. For solar-type stars, \dabb therefore represents the best simultaneous activity proxy known to date.
\end{abstract}




\section{Introduction}

A planet causes the radial velocity (RV) of its host star to change periodically over time. Yet, Doppler-like signals caused by the star itself, linked to the interplay between the evolving magnetic field and stellar surface convection, can drown out and mimic planetary signals. These manifestations of stellar magnetic activity represent a major obstacle to detecting planetary-induced RVs below 1 \ms \citep[see][]{Crass_2021}, with only very few measurements below this threshold \citep[e.g.][]{Faria_2022}. To date, Earth-like planets orbiting solar-type stars in the habitable zone are out of reach, as they produce RV signals with semi-amplitudes of the order of 10\cms. It is therefore essential to disentangle planetary and stellar RV components to obtain a clean planetary RV curve.

Stellar activity subsumes a range of phenomena including stellar magnetic cycles \citep{Lanza_2010,Costes_2021}, starspots \citep{Saar_1997,Desort_2007,Lagrange_2010}, faculae and plages \citep{Saar_1997,Saar_2003,Saar_2009,Meunier_2010a,Meunier_2010b}, meridional flows \citep{Meunier_2020}, granulation \citep{Dravins_1982,Dumusque_2011,Meunier_2015,Cegla_2019}, super-granulation \citep{Rieutord_2010,Rincon_2018,Meunier_2019}, and p-mode oscillations \citep{Mayor_2003,Medina_2018,Yu_2018,Chaplin_2019}.
These phenomena act on different timescales and have different impacts on the RVs.
An effect of particular importance is the suppression of convective blueshift \citep{Meunier_2010a}. For solar-type stars, the emission emanating from convective upflows dominates over the downflows and leads to a net blueshift of the stellar spectrum. However, this effect is modulated by the magnetic field inhibiting stellar surface convection \citep[e.g.][]{Hanslmeier_1991}. Since the magnetic field is spatially inhomogeneous, regions with suppressed convection rotate in and out of view as the star rotates, leading to a varying Doppler shift and variations in the shape of the absorption lines. In addition, the magnetic field evolves in time, and thus the overall effect also varies in time beyond the rotational timescale. 


%
The hemispherically averaged unsigned magnetic flux \bobs has been shown experimentally to be an excellent proxy for variations in solar RV \citep{Haywood_2016,Haywood_2022}. This finding is supported by analyses of Dopplergrams and magnetograms from the Michelson Doppler Imager \citep{Scherrer_1995} presented in \citet{Meunier_2010b}. They showed that suppression of convective blueshift is pronounced where the magnetic field is strong. In addition, simulations by \citet{Meunier_2010a} showed that the attenuation of the convective blueshift is indeed the dominant contributor to star-induced RV variations. The analyses in  \citet{Meunier_2010a} indicate that the attenuation of the convective blueshift leads to a long-term RV signal with an amplitude of about 8 \ms and thus impedes the detection of Earth twins orbiting solar-type stars. The photometrically induced RV variations due to bright active regions and dark starspots rotating in an out of view, on the other hand, partially cancel out and are of lesser concern.


The strength and evolution of stellar magnetic fields are challenging to measure, though.
The earliest measurements of the solar magnetic field date back to 1908, with \citet{Hale_1908} observing Zeeman splitting \citep{Zeeman_1897} in sunspot spectra.
Most of the methods that exist to date are either only applicable to highly active stars, require polarimetric data, measurements at infrared wavelengths or a combination of these \citep{Saar_1985,Valenti_1995,Johns-Krull_1999,Reiners_2006}.
An overview of magnetic field estimation methods is provided in Sections \ref{ss:polmeas} and \ref{ss:intensity_based_methods}.

In this study, we show how a proxy of \bobs can be derived from intensity spectra in the visible wavelength range of weakly active to moderately active stars. These stars have been, and continue to be, prime targets of RV surveys.
This work represents an extension of, and builds on, the Multi-Mask Least-Squares Deconvolution (MM-LSD) method\footnote{Available on github: \url{https://github.com/florian-lienhard/MM-LSD}.} presented in \citet{Lienhard_2022}. 
They analysed the performance of Least-Squares Deconvolution \citep [LSD;][]{Donati_1997} as a tool to extract RV information from spectra of FGK-type stars and the dependence of the measured RV on various parameters. For this purpose, \citet{Lienhard_2022} developed a pipeline that continuum-normalises deblazed echelle order spectra, partially corrects for telluric absorption lines, masks problematic wavelength regions, and finally extracts the RV via LSD. The preprocessing steps implemented in this pipeline and the convolution approach are reused in the present analysis.

In Section \ref{s:tb}, we explain the basics of Zeeman splitting, describe different magnetic field diagnostics based on this effect, as well as the in this context commonly used LSD method. In Section \ref{s:data}, we describe the data products used in this analysis. The proposed magnetic flux proxy is presented in Section \ref{s:methods}. We describe the application on HARPS-N solar spectra in Section \ref{s:application}. Lastly, we show and discuss our results in Section \ref{s:results}, and conclude in Section \ref{s:conclusion}.

\section{Theoretical background} \label{s:tb}
\subsection{Least-Squares Deconvolution}

The LSD method relies on physical information about the absorption lines, such as line depth and wavelength, to model the spectra at hand on the basis of few assumptions. The objective of using a simple model is to describe and model the bulk of the absorption lines, rather than to precisely model single lines.
In this way, one can tease out line information from all absorption lines and extract information that is otherwise hidden in the noise.
The LSD model for a spectrum is generated by convolving a common profile with a line list consisting of scaled delta functions at the rest wavelengths of the known absorption lines. The delta functions are scaled depending on what needs to be modelled. To model intensity spectra, for instance, we scale the delta functions by the expected relative depths of the relevant spectral absorption lines \citep[e.g.][]{Lienhard_2022}.

By applying least-squares fitting, the best-fitting common profile can be determined. Similarly to the Cross-Correlation Function \citep[CCF;][]{Baranne_1996,Pepe_2002}, this common profile represents the average shape of the absorption lines. Analogously to the CCF, one can extract the stellar RV from this common profile. Other applications include modelling Stokes $V$ spectra to estimate the stellar magnetic flux \citep[e.g.][]{Donati_1997}.

There are two main assumptions that LSD is based on. One is that absorption lines add up linearly. This assumption is valid for weak absorption lines only \citep[][]{Kochukhov_2010}. Secondly, LSD defines the common profile in velocity space. It is thus assumed that the absorption lines have the same shape in velocity space and only scale by a wavelength-specific factor.
This standard assumption in LSD \citep{Donati_1997,Kochukhov_2010,Lienhard_2022} is based on the fact that the conditions on the stellar surface are similar, all lines are rotationally broadened, and the dominant atomic absorbers for FGK-type stars, such as Fe, Ni, Cr, and Ti, all have similar atomic masses and hence similar thermal broadening. The width component due to thermal broadening is
\begin{equation}
    \Delta\lambda_{\text{T}} = 2 \frac{\lambda_0}{c} \sqrt{\frac{2kT}{m} ln( 2)},
\end{equation}
where $T$ is the plasma temperature, $c$ the speed of light, $k$ the Boltzmann constant, $m$ the atomic mass, and $\lambda_0$ the rest wavelength of the absorption line \citep[e.g.][]{Bellot_Rubio_2019}. Since this expression scales linearly with wavelength, the width contribution in the velocity domain remains constant\footnote{The conversion can be made with the standard approximation $\frac{\Delta \lambda}{\lambda} = \frac{\Delta v}{c}$ where $\Delta \lambda$ is the observed Doppler shift due to the source moving with velocity $\Delta v$ relative to the observer.}.
Lastly, micro- and macro-turbulent broadening are the same for all species.

The LSD method can therefore be a useful tool if we find the same general shape at the wavelengths of the absorption lines and we can model this shape as a profile fixed in velocity space scaling with factors that depend on the lines' physical properties.

\subsection{Zeeman effect} \label{ss:zeeman}
In this Section, we summarise the theoretical background of Zeeman splitting and the relevant equations describing its effect on absorption lines.
A magnetic field splits an absorption line involving a magnetically sensitive state into multiple absorption lines at slightly offset wavelengths. This effect is called Zeeman splitting and is due to the external magnetic field splitting an initially degenerate energy level with angular momentum $J$ into  $2J+1$ sublevels. The energy difference between the original degenerate state and the split components is proportional to the magnetic field strength and their respective magnetic quantum number $m$. Transitions involving these split states are therefore shifted in wavelength. For a review on Zeeman splitting in stellar spectra, see e.g. \citet{Reiners_2012,Stenflo_2013,Bellot_Rubio_2019}.

For electric dipole transitions, the selection rules allow for $\Delta~m$ = 0 and $\Delta~m$ = $\pm$ 1. The former produce the unshifted central $\pi$ component, while the latter produce the shifted $\sigma$ components. A simple triplet of lines is created when $J$ of one of the involved states is equal to 0 or if the Land\'{e} factors of both states are equal. More complicated patterns can still be treated as a triplet by computing an effective Land\'{e} factor $g_{\text{eff}}$ for the transition.
The polarisation of the individual components depends on the direction of the magnetic field relative to the observer. Crucially, the $\sigma$ components are circularly polarised with opposite orientation if the magnetic field vector points towards or away from the observer. Since the $\sigma$ components are shifted in wavelength with the shift proportional to the magnetic field strength, this generates a characteristic signal in Stokes $V$ and enables the estimation of the stellar magnetic flux from polarised spectra, as outlined in Section \ref{ss:polmeas}. For transverse fields, on the other hand, the three components are linearly polarised producing signals in Stokes $Q$ and $U$ \citep{Kochukhov_2011}.

The wavelength difference between the $\pi$ and the $\sigma$ components is equal to
\begin{equation} \label{eq:zeeman_lambda}
    \Delta \lambda = 4.67\times10^{-10} \: \lambda_0^2 g_{\text{eff}} B,
\end{equation}
which translates to a velocity shift of 
\begin{equation}\label{eq:zeeman_v}
    \Delta v = 1.4 \times 10^{-4} \: \lambda_0 g_{\text{eff}} B,
\end{equation}
with the rest frame wavelength $\lambda_0$ in \AA, $g_{\text{eff}}$ the dimensionless effective Land\'{e} factor, the magnetic field strength $B$ in kG, and the velocity shift in \kms. It follows that lines at longer wavelengths exhibit stronger Zeeman splitting. A field of 1 kG strength typically shifts an absorption line in the visible range by about 1\kms. This is orders of magnitudes larger than the typical RV shift due to a planet. However, the Zeeman signal manifests very differently in the absorption lines, as it leads to a varying shift of a varying fraction of the absorption lines rather than a velocity shift affecting the entire spectrum uniformly. Furthermore, only a tiny surface fraction (less than a few per cent for the Sun, as can be derived from \citealt{Milbourne_2021,Haywood_2022}) of weakly active stars is affected by such strong fields. The Zeeman signal is therefore washed out and intermixed with weaker splitting patterns. 
Since the width of a typical absorption line is much greater than 1\kms, Zeeman splitting generally leads to slightly broadened lines in the optical for FGK-type stars, rather than separated Zeeman triplets. 
For non-saturated lines Zeeman splitting does not alter the equivalent width. However, the splitting expands the saturation region of saturated absorption lines and thereby increases their equivalent widths. This effect is called Zeeman intensification \citep[e.g.][]{Saar_1992,Basri_1994,Kochukhov_2020}.

In the following two sections, we describe how polarised and unpolarised stellar spectra are affected differently by Zeeman splitting and how this relates to the techniques used to characterise magnetic fields.

\subsection{Polarimetric measurements} \label{ss:polmeas}

The 4th component of the Stokes vector, Stokes $V$, is defined as the difference between right and left-handed circular polarisation \citep[for a review see][]{Stenflo_2013}. The $\sigma$ components are oppositely circular-polarized when the magnetic field vector is parallel to the line of sight, as mentioned in Section \ref{ss:zeeman}. This leads to a characteristic signal in Stokes $V$ thus encoding the strength and orientation of the large-scale magnetic field. In the weak field regime, this information can be extracted from multiple lines simultaneously using LSD, as described in \citet{Donati_1997}.

The LSD method is used, for instance, to compute surface magnetic maps through Zeeman Doppler Imaging \citep[ZDI;][]{Semel_1989,Donati_zdi_1989,Kochukhov_2016}. 
ZDI capitalises on the fact that Stokes $V$ signatures of active regions are blueshifted as they first appear on the visible stellar hemisphere and then progressively shift towards longer wavelengths as the star rotates. Given a series of observations at different times, surface magnetic maps can be reconstructed by finding the maximum-entropy magnetic field geometry that produces the observed Stokes $V$ time series \citep[e.g.][]{Skilling_1984,Donati_2006,Folsom_2018}.

Methods based on Stokes $V$ permit the extraction of magnetic field diagnostics for rapidly rotating stars, but they have some inherent disadvantages. Mainly, Stokes $V$ signals from adjacent stellar surface regions of opposite polarity can cancel out if the respective polarised components are not sufficiently separated in wavelength \citep[e.g.][]{Saar_1988,Donati_1997,Reiners_2012}. This leads to an underestimation of the magnetic field strength.
Linear polarisation signals are much weaker, can be significantly affected by magneto-optical effects \citep{Landolfi_1982}, and are also subject to cancellation effects \citep{Saar_1988,Reiners_2012}.
For instance, \citet{Kochukhov_2011} find the linear polarisation signal to be 10 times weaker than the circular polarisation signal.
Lastly, taking polarimetric measurements requires additional equipment, poses technical challenges, and generally uses up more observation time to collect the same number of photons as compared to Stokes I measurements.

\subsection{Stokes I measurements} \label{ss:intensity_based_methods}

Extraction of magnetic field information from intensity spectra is fraught with technical complications but in principle has significant advantages over polarimetric methods. It is important to note that, conversely to Stokes $V$ measurements, intensity spectra are not affected by cancellation effects due to regions of opposite polarity.
Also, there are many high-resolution spectrographs designed for radial velocity studies on solar-type stars producing extensive time series of Stokes I spectra, but only few that provide polarimetric data. 
The capability to simultaneously measure the evolution of the instantaneous magnetic flux and the RV from intensity spectra is expected to lead to the discovery of smaller planets and improve mass measurements of known planets in the vast amount of existing and upcoming data.

Zeeman splitting measurements in the visible range are challenging since the line profile changes are very small in Sun-like stars and can be confused with other line broadening effects, such as thermal broadening \citep[see e.g.][]{Reiners_2012, Bellot_Rubio_2019}.
Since Zeeman splitting is proportional to the wavelength squared (see Eq. \ref{eq:zeeman_lambda}), many intensity-based methods therefore focus on extracting information from one suitable line in the infrared at very high spectral resolution. For instance, a few studies are based on the Mg I line at 12.32 $\mu$m \citep[e.g.][]{Brault_1983,Zirin_1989,Bruls_1995}.
The infrared domain poses instrumental problems, however, and is riddled with telluric absorption lines that can lead to a higher RV error \citep{Cunha_2014}. Water absorption lines are especially problematic as the precipitable water vapour content is spatially inhomogeneous and variable \citep[e.g.][]{Leet_2019,Cretignier_2021}.
Furthermore, most of the spectrographs designed for RV studies record stellar intensity spectra in the visible wavelength range. For solar-type stars, this wavelength range is optimal because there is a large number of absorption lines and the SNR is highest as the stellar flux peaks in the visible. For example, HARPS \citep{Mayor_2003} and HARPS-N \citep{Cosentino_2012} measure spectra from 383 to 690 nm, EXPRES from 390 to 780 nm \citep{Jurgenson_2016}, and ESPRESSO from  378.2 to 788.7 nm \citep{Pepe_2021}. Other instruments, such as CARMENES (520--960 and 960--1710 nm, \citealt{Quirrenbach_2016}) or NEID (380--930 nm, \citealt{Halverson_2016}) also record parts of the near-infrared spectrum. However, these instruments still do not reach the wavelength regime where cleanly split Zeeman diagnostic lines are found.
Hence, there is a need for a magnetic flux proxy for intensity spectra in the visible wavelength range.

There are some studies describing Stokes I magnetic field fitting techniques. \citet{Stenflo_1977} fit 402 unblended unpolarised Fe 1 absorption lines in the visible wavelength range observed at the solar disk centre. They fit the line widths at different depths and were able to estimate an upper limit for the average magnetic flux of 110 G for the Sun. 
\citet{Kochukhov_2020} have extracted magnetic flux estimates for Sun-like stars via Zeeman intensification. However, their error bars are much larger than the typical average magnetic flux variations of less than 1 G that are relevant for stellar activity mitigation. In this context, we present the first method able to produce sufficiently precise magnetic flux time series.

\section{Data} \label{s:data}

\subsection{VALD3}
To model the magnetic response of the absorption lines, we require the stellar absorption lines' wavelength, depth, and their effective Land\'{e} factor. This information can be retrieved from the Vienna Atomic Line Database (VALD3; \citealt{Ryabchikova_2015}).
Since we analyse solar spectra in this work, we set the stellar microturbulence to 1.1\kms, the effective temperature ($T_{\text{eff}}$) to 5833 K, and surface gravity ($\log g$) to 4.44, and the chemical composition to solar values. These stellar parameters were estimated as outlined in \citet{Lienhard_2022}, section 2.3. Our estimate for the solar effective temperature is marginally higher than the recommended value of 5772 K \citep{IAU_2016}. We kept the value that we derived from the HARPS-N spectra to keep the analysis consistent. We do not expect this to have any measurable impact on our results. We only included lines with relative depth greater than 0.2 to exclude very weak lines which are often affected by noise. Furthermore, we excluded all molecular absorption lines in the VALD3 list to have a more homogeneous set. This removes about 12 per cent of the lines in our list and very marginally improves our results. About 50 per cent of the remaining lines are due to Fe 1.

\subsection{HARPS-N} \label{s:hn}
HARPS-N is a pressure and temperature-stabilised cross-dispersed echelle spectrograph operational since 2012. It has a resolving power of R = 115,000 in the visible range from 383 to 690 nm over 69 spectral orders. In addition to the nightly observations, HARPS-N is outfitted with a solar telescope to record disk-integrated spectra of the Sun at 300-second cadence, and has been doing so for several hours on most days since 2015 \citep{Cosentino_2014,Dumusque_2015,Phillips_2016,Collier_Cameron_2019,Dumusque_2021}. 

For this study, we randomly selected one spectrum from each observing day contained in the set of three years of high-quality HARPS-N solar observations presented in \citet{Dumusque_2021}.\footnote{\url{https://dace.unige.ch/sun/}} We note that HARPS-N had a cryostat leak requiring periodic interventions. As a result, the spectra within 5 days from an intervention were excluded from this dataset because they can be affected by flux variations and are not representative of HARPS-N's usual performance.

In total, we selected 500 spectra. The exposure time was 300 seconds for each of them. The first spectrum was recorded on 29 July 2015, and the last spectrum on 18 May 2018. The airmass of the exposures ranges between 1 and 2.9, with the median being around 1.3. The minimum signal-to-noise ratio (SNR) at 550 nm is about 250, the maximum is around 460, and the median is equal to 380. These SNR values are very high compared to nightly HARPS-N observations with a typical SNR between 50 and 200. As we show in Section \ref{ss:snr_dependence}, our approach does not require a very high SNR.

For each spectrum, we extracted MM-LSD RVs as well as CCF RVs. Furthermore, the Data Reduction System (DRS) also computes several activity indices. For this study, we used the Full Width at Half Maximum (FWHM), contrast and bisector inverse slope (BIS) of the CCF. The RV, FWHM, and contrast values were corrected for effects of Solar System motions as detailed in \citet{Collier_Cameron_2019}. 
Furthermore, the \rhk index was computed directly from the HARPS-N spectra, yielding values between -5.03 and -4.96. This indicator quantifies the emission in the cores of the Ca II H (3968.47 \AA) and K (3933.66 \AA) spectral lines which is induced by magnetic activity. First, the S-index is computed standardly by weighing the emission within these bands with a triangular response function with width 1.09 \AA\ and dividing by the reference bands with a width of 20 \AA\ at 3900 and 4000 \AA\ \citep{GomesdaSilva_2011,Dumusque_2021}. The emission within the line cores was susceptible to contamination due to effects related to the cryostat leak. The leak led to the build-up of humidity over time, increasing the reflectivity in the detector and producing local flux variations called ghosts. The impact of these ghosts on the extracted S-index is corrected as described in \citet{Dumusque_2021}. The S-index is then converted to \rhk following \citet{Noyes_1984}.

\subsection{SDO}
To validate our results, we compare with data from the Helioseismic and Magnetic Imager (HMI) instrument onboard the Solar Dynamics Observatory \citep[SDO;][]{Schou_2012,Pesnell_2012,Couvidat_2016}.
HMI measures the line-of-sight magnetic flux through the magnetically sensitive Fe I line at 6173.3 \AA. We downloaded\footnote{\url{http://jsoc.stanford.edu/ajax/exportdata.html}} the 720 s magnetograms and intensitygrams using a six-hour cadence spanning our full HARPS-N time range. The absolute value of the SDO line-of-sight magnetic fluxes were intensity-weighted and summed over all pixels as outlined in \citet{Haywood_2016} to compute \bobs. 
To estimate the filling factors of active regions, we used the same thresholds as in \citet{Haywood_2016} and \citet{Milbourne_2019} to distinguish between faculae, sunspots, and quiet photosphere. More specifically, the magnetic field was assumed to be radial. Any pixel with foreshortening-corrected magnetic flux below 24 G was assumed to measure quiet photosphere. Pixels above this threshold were divided into sunspots and faculae using an intensity threshold of 0.89 times the mean pixel intensity corrected for limb-darkening as in \citet{Yeo_2013}. 
The filling factor and unsigned magnetic flux time series can alternatively be obtained using \texttt{SolAster}, presented in \citet{Ervin_2022}.

There is a minor time difference between the HARPS-N observations and the SDO magnetic flux measurements. This time difference is smaller than 6 hours (mean absolute difference: 2 hours) for all our measurements. Since \bobs and the RVs only marginally evolve over this timescale, the time difference only minimally influences our results. By interpolating the SDO data to the timestamps of the HARPS-N spectra, we achieve correlations about 0.005 higher than those reported in this analysis.

\subsection{Tapas}
The Transmissions of the AtmosPhere for AStromomical data database (TAPAS; \citealt{Bertaux_2014}) provides transmission spectra for the Earth's atmosphere. We use one arbitrarily chosen transmittance spectrum (La Palma Roque de los Muchachos Canarias Spain, 2018/3/30, 01:45:07, airmass 1.03) to identify and exclude spectral regions impacted by deep tellurics. More information is provided in Section \ref{ss:prepare_spectra} and in \citet{Lienhard_2022}.

\section{Extracting the unsigned magnetic flux} \label{s:methods}
In this Section, we first describe our model for the difference between a Zeeman-split and an unsplit absorption line and where this model is valid. We subsequently show how the difference between our spectra and a master spectrum can be fit using the LSD approach to extract a proxy for \bobs.

\subsection{Residual model} \label{ss:resid_model}
For simplicity, we assumed that the magnetic field strength distribution on the stellar surface can be captured by two components: the quiet surface and the active regions with magnetic field strength roughly three orders of magnitude higher. The exact values do not have to be fixed for the algorithm described below.
The intensity profile of a Zeeman-split line can then be modelled as in \cite{Title_1975}, \cite{Robinson_1980}, and \citet{Marcy_1982}:
\begin{equation} \label{eq.linemodel_title}
    I(\lambda) = \alpha\left(f(\lambda+\Delta\lambda)+f(\lambda-\Delta\lambda)\right) + \beta f(\lambda).
\end{equation}
The parameter $\alpha$ captures the emission from the shifted $\sigma$ components, $\beta$ quantifies the emission from unsplit lines and the unshifted $\pi$ component, $\Delta\lambda$ is the Zeeman shift proportional to the effective Land\'{e} factor and the magnetic field strength as in Eq. \ref{eq:zeeman_lambda}, and $f$ describes the shape of the line. Both values $\alpha$ and $\beta$ finally depend on the orientation of the magnetic field relative to the observer, as explained in Section \ref{ss:zeeman}.
The model in Eq. \ref{eq.linemodel_title} relies on the assumption that the $\sigma$ and the $\pi$ components have the same profile and do not interact. The validity of this assumption is based on the shifted $\sigma$ components having opposite polarity. This greatly simplifies the radiative-transfer problem, as it allows diagonalising the transfer matrix leading to non-interaction between the polarisation components \citep[e.g.][]{Stenflo_1984}. Note also that we assumed that the magnetic field strength does not vary radially within the active regions. We, therefore, assume that all lines are exposed to the same magnetic field independent of their formation height.

Assuming that the lines' equivalent widths remain the same, Eq. \ref{eq.linemodel_title} can be simplified:
\begin{equation}\label{eq.linemodel_gen}
    I(\lambda) = \alpha\left(f\left(\lambda+\Delta\lambda\right)+f\left(\lambda-\Delta\lambda\right)\right) + (1-2\alpha) f(\lambda).
\end{equation}
In the following, we assume that the ratio of $\pi$ to $\sigma$ components for the active regions remains about the same. In this case, the parameter $\alpha$ in Eq. \ref{eq.linemodel_gen} is directly proportional to the filling factor of active regions on the visible stellar hemisphere.\footnote{Assume the proportion of magnetic area: $r_{\text{active}}$, and non-magnetic area $1-r_{\text{active}}$ and ratio $r_{\text{s}}$ of $\sigma$ components. Then: $I(\lambda) = r_{\text{active}} (1-2r_{\text{s}})f(\lambda) + r_{\text{active}} r_{\text{s}} f(\lambda + \Delta\lambda) + r_{\text{active}} r_{\text{s}} f(\lambda - \Delta\lambda) + (1-r_{\text{active}})f(\lambda)$. This simplifies to $I(\lambda) = (1-2 r_{\text{active}} r_{\text{s}})f(\lambda) + r_{\text{active}} r_{\text{s}} f(\lambda + \Delta\lambda) + r_{\text{active}} r_{\text{s}} f(\lambda - \Delta\lambda)$, which is equivalent to our expression. $\alpha$ is thus equal to the filling factor of magnetic regions times a factor related to the distribution of the orientation of the magnetic field vectors.}
We note that the strength of the $\pi$ to $\sigma$ components for a given active region depends on the angle between the line of sight and the magnetic field vector \citep[e.g.][]{Seares_1913,Marcy_1982,Skumanich_2002}. By the Seares' relation, the intensity of one $\sigma$ component at a given position on the stellar surface is equal to
\begin{equation} \label{eq:seares1}
    I_{\sigma} = I_0 \frac{1+\cos^2\theta}{4},
\end{equation}
where $I_0$ is the total intensity of all three components and $\theta$ is the angle between the line-of-sight and the magnetic field vector. 
The strength of the $\pi$ component is then:
\begin{equation}
    I_{\pi} = I_0 \frac{\sin^2\theta}{2}.
\end{equation}
For disk-integrated spectra, \citet{Marcy_1982} assumed a radial field and estimated the average $\theta$ to 34$ ^{\circ}$.
Since we are mainly interested in the evolution of the magnetic flux, we do not need to estimate an average field-line to line-of-sight angle.
However, we assume that the active regions are homogeneously distributed such that in the disk-integrated spectra the ratio remains about constant. This means that the factor capturing the disk-averaged value of $0.25(1-\cos^2\theta)$ is absorbed in $\alpha$.

Lastly, we assume that the line profiles are all Gaussian. With few exceptions, this is a valid assumption for optical absorption lines of main-sequence FGK-type stars given our resolution and precision.
The intensity profile of an absorption line with line depth $d$, central wavelength $\mu$, and width $\sigma_{\lambda}$ is then
\begin{equation}\label{eq.linemodel_nomag}
    I_{\text{non-magnetic}}(\lambda) = d e^{-\frac{(\lambda-\mu)^2}{2\sigma_{\lambda}^2}}
\end{equation}
in the absence of magnetic flux. 
The line profile emerging from the active region is equivalently:
\begin{equation} \label{eq:mag}
    I_{\text{magnetic}}(\lambda) = d\alpha e^{-\frac{(\lambda \xi_{\text{R}}-\mu)^2}{2\sigma_{\lambda}^2}}+d\alpha e^{-\frac{(\lambda \xi_{\text{L}}-\mu)^2}{2\sigma_{\lambda}^2}} + d(1-2\alpha) e^{-\frac{(\lambda-\mu)^2}{2\sigma_{\lambda}^2}},
\end{equation}
where $\xi_{\text{L}}$ is equal to $1+\epsilon$ and  $\xi_{\text{R}}$ is equal to $1-\epsilon$ with $\epsilon$ representing the Zeeman-induced velocity shift (see Eq. \ref{eq:zeeman_v}) divided by the speed of light $c$:
\begin{equation}
    \epsilon = \frac{1.4 \times 10^{-4} \lambda_0 g_{\text{eff}} B}{c}.
\end{equation}
Subtracting the magnetic line from the non-magnetic line, we get
\begin{align}
    I_{\text{diff}}(\lambda) &= d\alpha (2 e^{-\frac{(\lambda-\mu)^2}{2\sigma_{\lambda}^2}}- e^{-\frac{(\lambda\xi_{\text{R}}-\mu)^2}{2\sigma_{\lambda}^2}}- e^{-\frac{(\lambda\xi_{\text{L}}-\mu)^2}{2\sigma_{\lambda}^2}})\label{eq:fdiff9}\\
    &= d\alpha  \left(g(\lambda;\xi_{\text{L}}) + g(\lambda;\xi_{\text{R}})\right) \label{eq:fdiff}.
\end{align}
with
\begin{equation} \label{eq:diff}
    g(\lambda;\xi) = e^{-\frac{(\lambda-\mu)^2}{2\sigma_{\lambda}^2}}-e^{-\frac{(\xi \lambda-\mu)^2}{2\sigma_{\lambda}^2}}
\end{equation}
The expression in Eq. \ref{eq:diff} is equal to the difference between two Gaussian absorption lines that are shifted relative to each other. It can be decomposed in terms of the Hermite-Gaussian polynomials as in \cite{Holzer_2020}:

\begin{equation} \label{eq:hgrv}
    g(\lambda;\xi) =\sum_{n=0}^{\infty}C_n(\xi)\psi_n(\lambda;\mu,\sigma_{\lambda}),
\end{equation}
where $\psi_n$ are the Hermite-Gaussian functions defined as in Eq. \ref{Aeq:hg} and $C_n$ are coefficients (Eq. \ref{Aeq:coeff0} and \ref{Aeq:coeffn}) that depend on the shift of the two Gaussians relative to each other. For more details, we refer to Appendix \ref{app:expansion} or \citet{Holzer_2020}.
By Taylor-expanding the relevant coefficients in $\epsilon$, a simple expression emerges, as the terms that are odd in $\epsilon$ cancel out due to the symmetry of Zeeman splitting. The odd components $C_1 \psi_{1}$ and $C_3 \psi_{3}$ can be neglected for the same reason. Thus, we get the following convenient expression:
\vspace{1cm}
\begin{equation}
    I_{\text{diff}}(\lambda) = 2d\alpha\epsilon^2\left(\frac{\mu^2}{2\sigma_{\lambda}^2} +\frac{3}{8} - \left(1+\frac{\mu^2}{2\sigma_{\lambda}^2}\right) \left(\frac{\lambda-\mu}{\sigma_{\lambda}}\right)^2\right)e^{-\frac{(\lambda-\mu)^{2}}{2 \sigma_{\lambda}^{2}}}.
\end{equation}
The parameter $\sigma_{\lambda}$ denotes the standard deviation of the Gaussian absorption line profile in wavelength space. We can translate this to the velocity space and assume that the width of all lines in this space is about the same, as in LSD. $\frac{\mu}{\sigma_{\lambda}}$ then translates to $\frac{c}{\sigma_v}$ and $\frac{\lambda-\mu}{\sigma_{\lambda}}$ to $\frac{v-v_0}{\sigma_{v}}$ with $v_0$ the absolute radial velocity of the star:
\begin{equation} \label{eq:idiff_v}
    I_{\text{diff}}(v) = 2d\alpha\epsilon^2\left(\frac{c^2}{2\sigma_{v}^2} +\frac{3}{8} - \left(1+\frac{c^2}{2\sigma_{v}^2}\right) \left(\frac{v-v_0}{\sigma_{v}}\right)^2\right)e^{-\frac{(v-v_0)^{2}}{2 \sigma_{v}^{2}}}.
\end{equation}
Note that keeping the dominant components of Eq.\ref{eq:idiff_v}, we recover the shape of the second derivative of the absorption line (cf. Eq. \ref{eq:idiff_v_approx}).

\subsection{Range of validity} \label{ss:validity}
\subsubsection{Residual approximation}
We included only the dominant quadratic terms of the Taylor expansion in our estimation of the residual shape in Eq. \ref{eq:idiff_v}. 
Quartic and later terms only become relevant if $\frac{\mu}{\sigma_{\lambda}} > \frac{1}{\epsilon}$, i.e. if $\frac{\sigma_{v}}{c} < \epsilon$. 
Our approximation is valid if the width ($\sigma_v$) of the absorption lines in velocity space is larger than the Zeeman shift $1.4 \times 10^{-4} \lambda_0 g_{\text{eff}} B$.
While the condition above resembles the weak-field condition, it only refers to the validity of the Taylor expansion.

Note that the expression in Eq. \ref{eq:idiff_v} is equally valid for local as well as for disk-integrated spectra, given our assumptions. 
The reason for this lies in the fact that we model absorption lines as Gaussian functions and rotational and macroturbulent broadening are well described by a convolution with a Gaussian kernel \citep{Takeda_2017,Sheminova_2019}.
A Gaussian $N(\mu,\sigma^2)$ convolved with a Gaussian kernel $N(0,\sigma_b^2)$ results in $N(\mu,\sigma^2+\sigma_b^2)$ and therefore remains Gaussian in shape. 
Our residual shape is expressed as a sum of Gaussian functions. To translate this residual expression from the local to the disk-integrated spectra, it is convolved with a Gaussian kernel.
Since the convolution is distributive, the Gaussian functions in the sum can be individually convolved with the broadening kernel.
Therefore, the residual shape is equally a sum of Gaussian functions for the local and disk-integrated spectra. The only difference is the line width and line depth. Since we assumed constant $\sigma_v$ for the line profiles, the line widths are equally constant for the local profiles. By convolving with a Gaussian kernel, the line depths change by a multiplicative factor. Therefore, the extracted residual shape amplitudes scale by a factor that depends on the broadening kernels.
A multiplicative factor is of no concern for the purpose of RV detrending, however, and can remain unaccounted for. For our purposes, we investigate disk-integrated residual spectra, and therefore we need to ensure the Taylor expansion is a good fit to those spectra. Therefore, we set $\sigma_v$ to the line width measured from these spectra.

Our approximation is generally valid for absorption lines of weakly active FGK-type stars in the optical wavelength range. For the Sun, $\sigma$ is about 3\kms and a 1 kG magnetic field produces a typical Zeeman shift of about 1\kms, which therefore comfortably lies within our range of validity. 
The typical magnetic field strength of plage regions on the Sun, however, is about 1.5 kG with some areas reaching up to 2 kG \citep[e.g.][]{Rueedi_1992,MartinezPillet_1997,Buehler_2015}.
Similarly, magnetic field concentration in the solar network reach the same magnetic flux strengths \citep{Buehler_2019}.
In the following, we investigate the behaviour of the residual profile for a magnetically sensitive line for a magnetic field strength of 1 and 2 kG.

\begin{figure}
    \centering
    \includegraphics[width=\columnwidth]{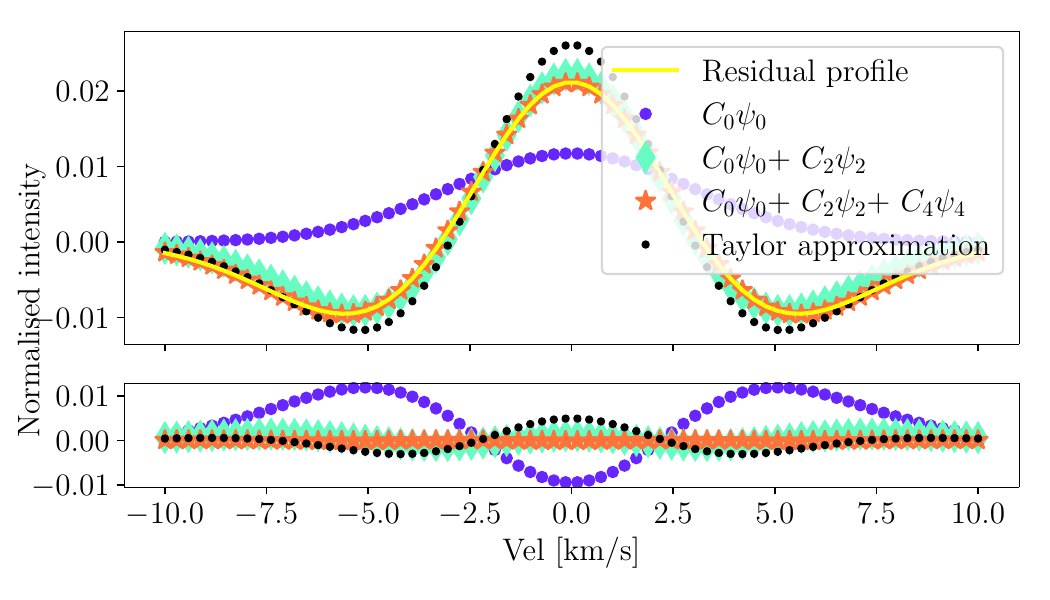}    \caption{\textit{Upper panel}: Residual profile emerging from subtracting a Zeeman-split absorption line exposed to a 2 kG field from a non-split line (yellow) and approximations thereof. \textit{Lower panel}: Difference between the residual profile and the approximations.}
    \label{fig:hg_components}
\end{figure}

\begin{figure}
    \centering
    \includegraphics[width=\columnwidth]{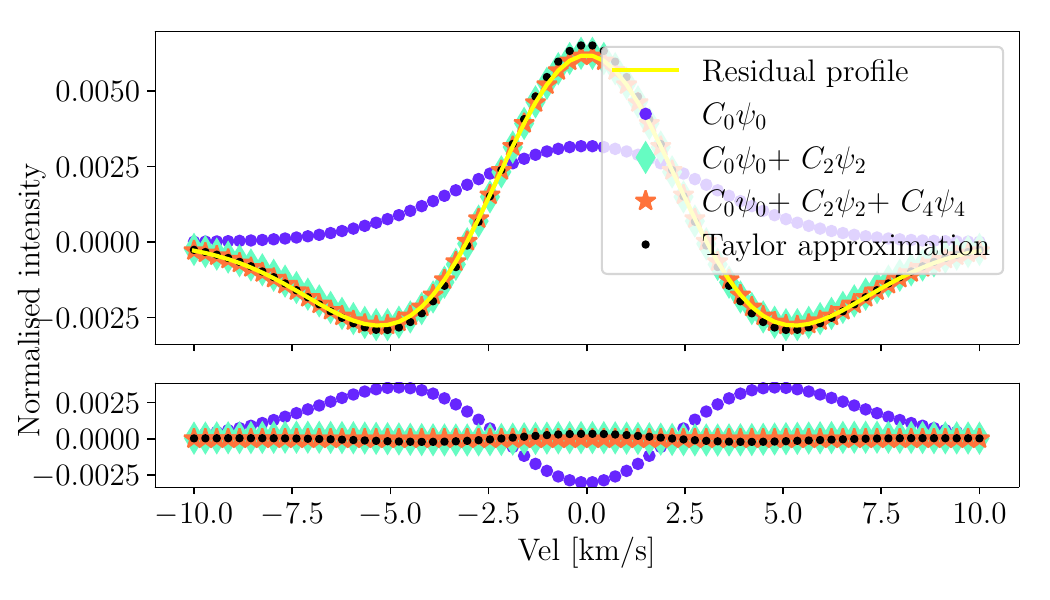}
    \caption{\textit{Upper panel}: Difference between a Zeeman-split absorption line exposed to a 1 kG field and a non-split line (yellow) and approximations thereof, as in Fig. \ref{fig:hg_components}. \textit{Lower panel}: Difference between the residual profile and the approximations.}
    \label{fig:hg_components1G}
\end{figure}

In Fig. \ref{fig:hg_components}, we show the residual profile computed from the original Gaussian expression in Eq. \ref{eq:fdiff9} in yellow, as well as the dominant Hermite-Gaussian components and our approximation as in Eq. \ref{eq:idiff_v}. For this example, we chose an absorption line at 5000 \AA, with $\sigma$ equal to 3\kms, relative depth 0.3, $g_{\text{eff}}$ of 2, $\alpha$ set to 0.1 and the magnetic field strength in the active region to 2 kG. This results in a Zeeman shift of 2.8\kms which is just within our validity range of 3\kms in this case. For comparison, we display the approximations for a 1 kG field keeping the other line parameters the same in Fig. \ref{fig:hg_components1G}. Since the residual profiles scale with $B^2$, the amplitude is reduced by a factor of 4 in Fig. \ref{fig:hg_components1G}. Note that only small fractions of the solar surface are affected by such strong fields \citep[e.g.][]{Haywood_2016} and only about 3 per cent of the included lines have an effective Land\'{e} factor greater than 2.

For both cases, $C_{0} \psi_{0} + C_{2} \psi_{2}$ is already very close to the numerical solution. The difference between our Taylor approximation as in Eq. \ref{eq:idiff_v} and the Hermite-Gaussian approximations $C_{0} \psi_{0} + C_{2} \psi_{2}$ and $C_{0} \psi_{0} + C_{2} \psi_{2} + C_{4} \psi_{4}$ lies mainly in the quartic components. Given that the example shown in Fig. \ref{fig:hg_components} treats a case that is at the very edge of the validity range and given the noise present in the spectra, including quartic components is not warranted. Also, the quartic components break the linearity of the problem that allows us to use the LSD approach in Section \ref{ss:conv_model}.

Apart from the question of when our Taylor expression in Eq. \ref{eq:idiff_v} is valid, we also need to consider when the underlying parametrisation in Eq. \ref{eq:mag} is valid on the local level where the lines are narrower. The parametrisation in Eq. \ref{eq:mag} is based on the weak-field approximation which breaks down for absorption lines with very high Land\'e factors in active regions with high magnetic field strengths \citep{Jefferies_1989,Lehmann_2015}. The violation of this condition may have a larger impact on the extraction of \dabb for more active stars but overall holds for solar-type stars. For now, we thus recommend applying our model to the spectra of solar-type stars.

\subsubsection{Gaussian absorption lines}
In the derivation, we assumed a line is well characterised by a Gaussian profile. Deviations from Gaussian profiles can arise due to pressure broadening leading to more prominent line wings. 
These absorption lines are generally better characterised by Voigt profiles. The residual profile emerging from these lines is still symmetric and broadly follows our approximation such that Zeeman-induced variability can still be captured.
Another factor leading to deviations from the Gaussian line shape is stellar surface convection \citep[see e.g.][]{Gray_2005}. The latter produces slightly asymmetric absorption lines in disk-integrated spectra \citep{Gray_2005,Cegla_2019}. 
More specifically, the asymmetry is due to granulation. As we observe a star, we record both the blue-shifted light emitted from upwards-flowing hot matter in the granules and the redshifted light emitted from the downwards-flowing cooler matter in the surrounding intergranular lanes. Averaged over the stellar hemisphere, this leads to C-shaped bisectors.
A bisector is defined as the line connecting the mid-points of each horizontal segment of an absorption line \citep[see e.g.][p.297]{Gray_2005}.
For our purposes, the variation of the line shapes in time is more important than the line shapes themselves. To investigate the residual shapes caused by bisector variations, we analysed the CCF variations of HD 166435\footnote{The spectra are available on the ELODIE archive: \url{http://atlas.obs-hp.fr/elodie/}.} recorded with the ELODIE spectrograph \citep{Baranne_1996}. This star is known to have large bisector variations that mimic the presence of a planet \citep{Queloz_2001}. The CCF variations are unsurprisingly not symmetric relative to the line centre but anti-symmetric instead. This result is expected because the line cores shift RV-like leading to shapes that are well characterised by the odd first Hermite-Gaussian function but are not expected to interfere when fitting an even function that also naturally places most of the statistical weight on the line centres. We leave the detailed analysis of strong line shape variations on our residual approach to future work.

In our derivation, we assumed that the equivalent width is conserved. This is not the case for saturated lines displaying Zeeman intensification. Saturation is an issue when absorption lines are to be modelled. However, as we model residual spectra and the Zeeman-induced line variations are very small, the impact of saturation is reduced. Nevertheless, we expect improved results if saturation can be incorporated into the residual model.

Blended lines are also known to pose problems to magnetic field extraction techniques. For this reason, unblended lines are selected in most works \citep[e.g.][]{Stenflo_1977,Giampapa_1983,Saar_1988}. As we include a large number of absorption lines, we expect the blend effects to be strongly diluted. We did, however, remove heavily blended regions and heavily saturated lines, as outlined in Section \ref{ss:prepare_spectra}.

\subsubsection{Zeeman triplets}
We modelled the residual profiles as triplets, despite most absorption lines not being simple triplets.
As explained in Section \ref{ss:zeeman}, an energy level with angular momentum $J$ splits into $2J + 1$ sublevels, which is of course the case for both states involved in a transition. If both states have non-zero angular momentum and different Land\'{e} factors, we see multiple $\sigma$ components at different wavelength shifts. 
The effective Land\'{e} factor is defined as the factor that captures the shift of the centre of gravity of the red-shifted $\sigma$ components \citep{Landi_1982}.

Since the $\sigma$ components are defined to originate from the transitions with $\Delta~m$ = $\pm$ 1, they come in pairs of a red-shifted and a blue-shifted component, as long as there is a state to populate, with the $\pi$ component in between. However, triplets can also be shifted relative to the non-magnetic case. This shift is proportional to the difference between the Land\'{e} factors of the upper and the lower state and produces a distribution of $\pi$ components that is symmetric relative to the non-magnetic transition wavelength.

As long as the Land\'{e} factors of both involved states are very similar, we can treat them as a superposition of unshifted triplets with different strengths.
Therefore, these non-triplets produce a superposition of residual profiles with the same shape but different amplitudes.
To first order, the effective Land\'{e} factor squared captures the amplitude of this residual profile. Thus, we can also model non-triplets with the triplet model by using the effective Land\'{e} factor, as done in this study.

We investigated the dependence of our results on the non-triplet transitions. We could remove up to 60 per cent of the absorption lines without noticeably deteriorating the results. This cut corresponds to excluding all absorption lines for which the Land\'{e} factors of the upper and the lower state differ by more than 0.1. This shows that, on the one hand, non-triplets do not interfere with our extraction. On the other hand, it shows that the magnetic information in the non-triplet lines is not yet fully harnessed.

\subsection{Convolution model} \label{ss:conv_model}
In this Section, we show how the line model can be applied on multiple lines simultaneously to boost the residual signal.
For this, we assume that line residuals add up linearly, which allows us to model the residual spectrum as a convolution using the residual profile
\begin{equation} \label{eq:idiff_v_r}
    R(v) = \left(\frac{c^2}{2\sigma_{v}^2} +\frac{3}{8} - \left(1+\frac{c^2}{2\sigma_{v}^2}\right) \left(\frac{(v-v_0)}{\sigma_{v}}\right)^2\right)e^{-\frac{(v-v_0)^{2}}{2 \sigma_{v}^{2}}}.
\end{equation}
scaled by the expected amplitude of the signal, $2d\alpha\epsilon^2$.
Note that the amplitude of the signal can be split into a line-specific component 
\begin{equation}
    2d\left(\frac{1.4\times 10^{-4}\lambda g_{\text{eff}}}{c}\right)^2
\end{equation}
multiplied by a general part \abb for all absorption lines.
We now have a profile that is constant in velocity space and only scales with a line-specific amplitude and a multiplicative factor. Therefore, we can use the LSD approach and model the convolution via matrix multiplication. For this, we need the following definitions:

\begin{itemize}
    \item $\lambda_i$: wavelength of pixel i.
    \item $\lambda_l$: central rest frame wavelength of absorption line $l$.
    \item $d_l$: depth of absorption line $l$.
    \item $w_l$: line-specific amplitude 2$d_l (1.4\times 10^{-4} \lambda_l g_{\text{eff}}/c)^2$.
    \item $v_j$: velocity grid point $j$ of residual profile $R$.
    \item $v_{il} = c \frac{\lambda_i - \lambda_l }{\lambda_l}$: radial velocity $v_{il}$ which shifts $\lambda_l$ to $\lambda_i$.
\end{itemize}

With these definitions, the convolution can be expressed as the matrix multiplication 
\begin{equation}
    I_{\text{diff}_\text{model}} = \alpha B^2 \textbf{M} R
\end{equation}
using
\begin{equation}
    \textbf{M}_{ij} = \sum_{l} w_l \: \Lambda\left(\frac{ v_j - v_{il} }{\Delta v}  \right),
\end{equation}
where $\Delta v$ is the velocity increment, and $\Lambda$ is defined as in \citet{Lienhard_2022}:
\begin{equation}
    \Lambda \left( x \right) = \begin{cases}
    0 & \left|x\right| \geq 1 \\
    1 - \left|x\right| & \left|x\right| < 1.
    \end{cases}
\end{equation}

To find the best-fitting model to the data, we compute the value of \abb that minimises
\begin{equation} \label{eq:minimisation}
\chi ^2 =  (I_{\text{diff}_\text{measured}} - \alpha B^2\textbf{M} R)^\intercal \textbf{S} (I_{\text{diff}_\text{measured}} - \alpha B^2\textbf{M} R),
\end{equation}
where the matrix $S$ contains the inverse squared flux uncertainties on the diagonal.

\subsection{Non-magnetic profile}

Above we estimated the residual profile arising from subtracting a non-magnetic line from a line affected by Zeeman splitting. However, we do not have an exact model for the unsplit line because the magnetic field is non-zero even in the quiet regions of the solar photosphere and we measure disk-integrated spectra. \cite{Robinson_1980} chose lines with low Land\'{e} factor as a model for the unsplit lines. However, for the line comparison to be meaningful, the formation heights of the split and the unsplit lines must be similar and the equivalent width must be comparable \citep{Robinson_1980}. Alternatively, the spectra of another magnetically quiet star of the same spectral type may be used as a comparison spectrum as in \citet{Giampapa_1983}. However, this approach comes with some complications, such as accounting for residual differences between the stars and estimating the magnetic flux of the quiet star \citep{Saar_1988}.

Another option consists in comparing the absorption lines to a reference spectrum of the same star \citep{Saar_1988,Thompson_2017,Thompson_2020}. 
This means that each absorption line is compared to a line with the same formation height and practically identical equivalent width.
Instead of choosing one spectrum as the reference, we compare each spectrum to the average spectrum computed from a selection of spectra. This reduces the extent of spurious variations due to telluric absorption lines, the continuum correction, and photon noise in the master profile leading to a cleaner comparison profile. 
We found this averaging procedure to be crucial as it significantly suppresses the photon noise in the template and therefore also reduces the scatter in the extracted values of \abb.

To generate the master spectrum, we stack 100 normalised solar spectra in the barycentric reference frame and fit a univariate spline. The exact selection of solar spectra is negligible for this process.
The residual profile evolves smoothly across the full velocity grid spanning 20 \kms with the two minima being about 10 \kms apart. This is orders of magnitudes larger than the expected RV shift due to planets. The extracted amplitude of the residual shape is thus not affected by planet-induced RV variations and $v_0$ in Eq. \ref{eq:idiff_v_r} can be set constant. Furthermore, Doppler-shift-induced residuals have a very different shape as compared to broadening-induced residuals. The extracted \abb is therefore largely independent of the presence of planets. Nevertheless, we shift the master profile to the RV of the individual spectra to match the derivation of the residual profiles.

Since we compare the absorption lines to their individual average profiles, we extract the change in magnetic flux strength \dabb, rather than the absolute $B$ or \abb:
\begin{align}
    I_{\text{diff}_{\text{measured}}} &= I_{\text{master}} - I_{i}\\
    &=(I_{\text{non-magnetic}} - I_{i}) - (I_{\text{non-magnetic}}-I_{\text{master}})\\
    &=I_{\text{diff}}(\alpha_{i},\epsilon_{i})-I_{\text{diff}}(\alpha_{\text{master}},\epsilon_{\text{master}}) \label{eq:splinediff1}\\
    &=(\alpha_{i} B_{i}^2-\alpha_{\text{master}} B_{\text{master}}^2) M\\
    &\propto \Delta ~\alpha B^2
\end{align}
This expression is valid under the assumption that what actually evolves is the filling factor of active regions on the visible stellar hemisphere rather than the unsigned magnetic flux within the magnetically active regions. 
To verify this, consider a triplet as in Eq. \ref{eq:mag}. Keeping all parameters the same but varying only $\alpha$, we see that the mean of any number of triplets can also be modelled as a triplet with well-defined parameters $\alpha$ and $\epsilon$. Within our toy model, the difference between the master profile and a non-magnetic line can therefore be modelled with Eq. \ref{eq:fdiff}, as we did in Eq. \ref{eq:splinediff1}.

However, our method does not rely critically on this assumption. Injection-recovery tests show that we can vary $B$ within the defined validity range (cf. Section \ref{ss:validity}) and keep $\alpha$ constant and still recover \dabb with a mere constant offset which does not interfere with the linear correlations. Note that we measure \dabb relative to the master profile which means that the difference in \dabb between two spectra is equal to the difference between their respective \abb and thus an overall offset is not worrisome.

Measuring the absolute evolution of the magnetic flux requires two measurements to calibrate \abb and determine the offset. These measurements can be carried out with the established magnetic field estimation approaches. The technique presented in this study consequently also provides a gateway to get precise and cost-effective absolute unsigned magnetic flux time series. 

Lastly, we assumed in Eq. \ref{eq:mag} that the magnetic field strength is radially constant within the active and quiet regions. This assumption is not critical for our application because we are only interested in how the residual profiles evolve. In fact, this assumption is of no concern if the magnetic flux within both regions is constant and only the filling factor of active regions evolves in time.

\subsubsection{Direct \abb extraction from the spectra}
We investigated applying an extended version of LSD modelling the absorption lines as triplets to the spectra themselves rather than the residual spectra. However, the combined problem consisting of the problematic LSD-intrinsic line addition assumptions and imperfect absorption line depths represented a Gordian Knot that we could only cut by applying the weak-field splitting approach on the residual spectra, at the expense of getting the variation in \abb rather than \abb itself. As mentioned, this does not impact RV mitigation and can be overcome by doing two calibration measurements.

\subsection{Relation to hemispherically averaged unsigned magnetic flux \bobs} \label{ss:rel_to_B}

As shown in the preceding sections, we extract \dabb from our spectra. 
SDO data shows that the changes in \bobs are mainly driven by the variation of the filling factor of active regions rather than the magnetic flux within those regions. In fact, in our SDO test dataset containing one SDO observation at 6-hour cadence from July 2015 to September 2021, we find the overall filling factor of active regions to correlate almost perfectly with \bobs (Pearson correlation coefficient: 0.99). Similarly high correlations between the filling factor and \bobs were found for the data set analysed in \citet{Ervin_2022}. For the Sun, this correlation is mainly driven by faculae (Pearson correlation coefficient: 0.98) as the Sun is a faculae-dominated star.
The correlation of the filling factor with the emission reversal in the Ca II H and K lines, as measured by \rhk, and the correlation between \rhk and the average magnetic flux was shown in \citet{Meunier_2018} and is also discussed in \citet{Haywood_2022}.

Assuming no magnetic field in the quiet regions and constant magnetic field strength in the active regions, the only factor that evolves in time is $\alpha$. This factor includes variations in the ratio of transverse to longitudinal field components, which we assume to be negligible in the disk-averaged spectra. In this case, $\alpha$ is directly proportional to the filling factor of active regions. Since we assumed a constant magnetic field strength in the active regions, \abb is directly proportional to the filling factor of active regions multiplied by their magnetic field strength. The latter is equal to \bobs in this idealised two-component model. 

A two-component model is supported by evidence that plages exhibit fairly tight distributions of the magnetic field strength \citep[e.g.][]{Rueedi_1992,MartinezPillet_1997,Buehler_2015}. This is due to the efficient concentration of the small-scale magnetic fields in flux tubes through the convective collapse mechanism \citep{Parker_1978,Spruit_1979}. 
For sunspots, there is a wider distribution of magnetic field strengths between spots of different sizes and within the individual spots themselves. The peak magnetic field strength is found in the umbra reaching 2000 -- 3700 G and decreases towards the periphery of the spot to 700 -- 1000 G \citep{Solanki_2003}. Since all of these components contribute to the amplitude of the residual signal described in this study, we expect the spread of the magnetic field strengths to impact the scaling of \dabb with \bobs for spot-dominated stars. Therefore, we expect the direct proportionality between \dabb and \bobs to hold for less active stars, as they are expected to be plage-dominated \citep[e.g.][]{Radick_2018}.

\subsection{Comparison of methodology to other techniques}

\citet{Skumanich_2002} applied Principal Component Analysis (PCA) on intensity spectra. They found the first component to be equal to the first derivative of the profiles and their score to correlate with the RVs. This finding is in perfect agreement with the results from \citet{Holzer_2020} who find that the difference between two shifted Gaussians is well-modelled by one profile with the shape of the first Hermite-Gaussian polynomial scaled with the stellar RV. \citet{Skumanich_2002} also find the second derivative to scale with the filling factor multiplied by the magnetic field strength squared. This agrees with our findings that the Zeeman signature can also be modelled by one profile that scales with $\alpha$ and the squared unsigned magnetic field strength of the active regions.
This result is also apparent from Eq. 10 in \citet{Stenflo_2013}. 

For the range of validity defined in \ref{ss:validity}, the difference between a line and a line broadened by a factor $1+r$ while preserving the equivalent width can be expressed as
\begin{equation} \label{eq:sigma_broaden}
    d e^{-\frac{(v-v_0)^2}{2\sigma_{v}^2}}-\frac{d}{1+r} e^{-\frac{(v-v_0)^2}{2(\sigma_{v} (1+r))^2}}
\end{equation}
By Taylor expanding this expression for small width variations, i.e. for $r$ near 0, it can be seen that this residual profile is practically identical to the expression in Eq. \ref{eq:idiff_v} for $r$ equal $\alpha\epsilon^2 \frac{c^2}{\sigma_{v}^2}$. Since the amplitude of the residual signal in Eq. \ref{eq:idiff_v} scales linearly with \abb, the residual profile in Eq. \ref{eq:sigma_broaden} scales linearly with $r$ for small width variations. The derivation is shown in detail in Appendix \ref{broadening_residual_profile}.
The emergent residual profile is therefore not unique to Zeeman splitting. {This also means we can model other broadening mechanisms by adding the same residual profile scaled by the appropriate factor.} Also, if Zeeman splitting is modelled as a line-broadening effect, we expect to get the same scaling behaviour.
Indeed, \citet{Stenflo_1977} fit the absorption line width of 402 unblended Fe 1 absorption lines as a polynomial expression with the magnetic factor scaling with $B^2$.

\citet{Lehmann_2015} applied a Principal Component Analysis (PCA) approach on the moderately active star $\epsilon$ Eridani. They find one eigenprofile in good agreement with the second derivative of the line profiles and extract $B^2$ via the respective principal component score. We do not extract the magnetic flux profiles from the spectra themselves. Instead, we derive the shape and scaling behaviour from our weak-field triplet splitting model and thus assume the residual profiles to be known \textit{a priori}. As also pointed out in \citet{Lehmann_2015} and following directly from the Hermite-Gaussian expansion, the purely RV-induced residual variations are orthogonal to the Zeeman splitting induced shape variations. These components, therefore, interfere negligibly which makes the modelling of the RV effect in the residuals unnecessary. \citet{Lehmann_2015} choose 30 spectral lines with Land\'{e} factor greater than 1.59. They generate a calibration mapping from the extracted $B^2$ values to the average magnetic flux by comparison with synthetic line profiles. Such a mapping is challenging for our technique because we include over 4000 absorption lines. 
The magnetic flux values of \citet{Lehmann_2015} vary by a few tens of Gauss from spectrum to spectrum, with the average magnetic flux being 186 G. This error ratio may inhibit stellar activity mitigation.

There is direct observational evidence for the existence of the features that we derive from Zeeman splitting in the present analysis. \citet{Thompson_2017} compared HARPS spectra of $\alpha$ Cen B to investigate the impact of stellar activity on absorption lines. Their data set spans a range of \rhk values from -5 to -4.82 and covers a sizeable fraction of $\alpha$ Cen B's activity cycle of about 8.1 years \citep{Ayres_2014}. This range of \rhk is larger than for the solar data used in this work implying a stronger variation of the magnetic flux $\alpha$B \citep{Schrijver_1989}.
They generated a low-activity stellar template by stacking the spectra recorded during a night in 2008 when $\alpha$ Cen B was most inactive within their data set. \citet{Thompson_2017} then divided the nightly stacked spectra from 2010, when $\alpha$ Cen B was more active, by this template to investigate the differences. By visual inspection, they found features that closely resemble the Zeeman-induced residual shapes derived in the present analysis. 
The fact that \citet{Thompson_2017} used division while we used subtraction does not significantly affect the morphological similarity. However, the Hermite-Gaussian approach we use only applies to residual spectra produced by subtraction.

By simulating specific absorption lines, \citet{Thompson_2017} deduced that magnetically active regions can produce the observed residual shapes. They furthermore found that the strength of the central component of the residual shape correlates with \rhk.
The residual shapes were also seen in a follow-up study in which \citet{Thompson_2020} stacked daily HARPS-N solar spectra to discover a number of the same features as in $\alpha$ Cen B. They also found that the strength of those features correlated well with \rhk and with the facular filling factor which agrees with our findings.
We do not visually see the Zeeman residual shapes, however, since we analyse individual spectra of the Sun. We can only deduce the magnetic flux proxy by combining the information from thousands of lines. This approach does make it possible to use this proxy as a simultaneous indicator for RV variations. As \citet{Thompson_2017} see a strong correlation for single lines, we expect K-dwarfs such as $\alpha$ Cen B to be exquisite targets for our multi-line approach too, as long as the extent of Zeeman splitting is within our range of validity and they are still plage-dominated.

\section{Application to HARPS-N solar spectra} \label{s:application}
In this Section, we describe how we applied the residual model to our data set of 500 HARPS-N solar spectra to extract \dabb. We subsequently outline the results from injection-recovery tests using the solar spectra and describe a combination of the magnetic field modelling with RV modelling.

\subsection{Preprocess spectra} \label{ss:prepare_spectra}

For the present analysis, we used the deblazed 2-dimensional echelle order spectra, their associated uncertainties, and barycentric wavelengths contained in the spectral files that we selected as described in Section \ref{s:hn}. We continuum normalised the spectra with RASSINE \cite{Cretignier_2020}, corrected for residual cryostat leak effects, and divided by a simple telluric model. Details can be found in \citet{Lienhard_2022}, section 4.

We excluded any wavelength range in the barycentric frame that is affected by a telluric line deeper than 1 per cent in any of the spectra. Such a strict threshold is warranted since we measure very small signals that could easily be distorted by telluric absorption lines \citep{Cunha_2014,UlmerMoll_2019}. 

\begin{figure}
    \centering
    \includegraphics[width=\columnwidth]{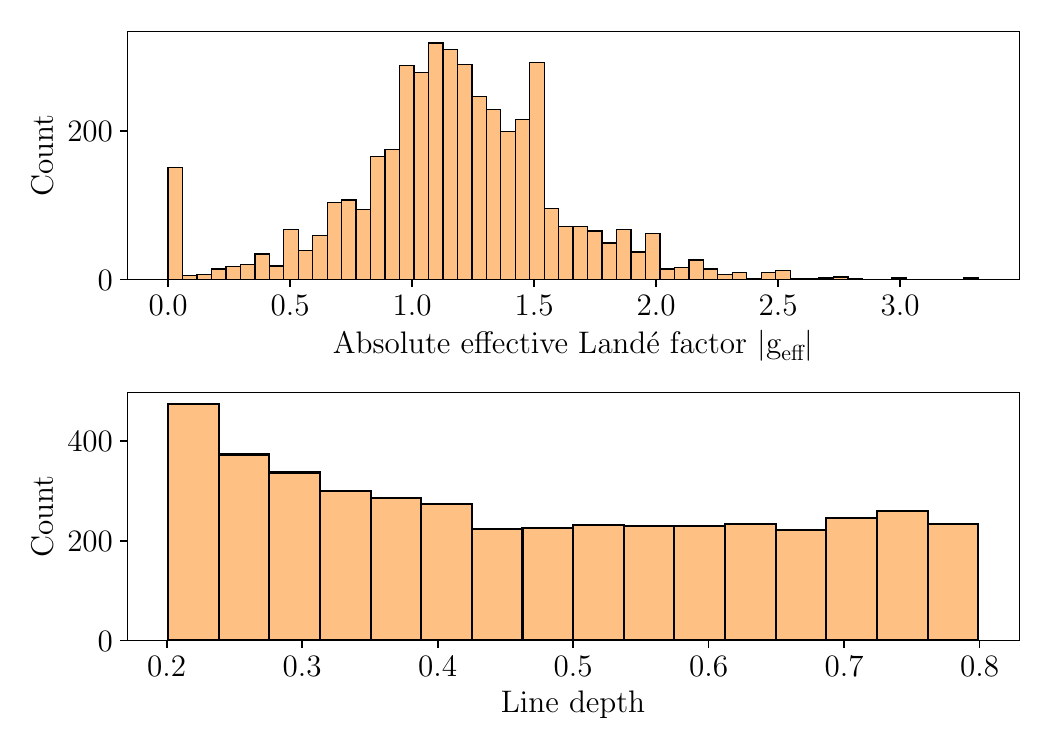}
    \caption{\textit{Top panel:} Histogram of the effective Land\'{e} factors of the included absorption lines. \textit{Bottom panel:} Histogram of the absorption lines depths from VALD3.}
    \label{fig:lines}
\end{figure}

In \citet{Lienhard_2022}, the spectra were modelled by convolving the best-fitting common profile, representing the average line profile, with a line list containing the wavelength and depth of the absorption lines.
For this, the velocity grid on which to evaluate the common profile had to be defined. We adopt the same velocity grid centred at the stellar RV of the first spectrum. The width of the velocity grid was set to 3 times the FWHM of the first common profile and the velocity increment to 0.82\kms. The latter is equal to the average velocity increment per physical pixel on the HARPS-N CCD. This results in a grid width of about 20\kms. Moreover, we excluded regions of the spectrum containing fluxes deviating by more than 0.5 in relative depth from the convolution model. This essentially removes lines that are heavily blended or absorption lines with inaccurately estimated depths in the VALD3 list. Furthermore, we included only atomic lines with relative depth greater than 0.2 as per the VALD3 list. We excluded all lines deeper than 0.9 and spectral regions near such absorption lines to avoid including heavily saturated absorption lines which are not well modelled by our residual model. This leaves us with 4636 absorption lines with mean effective Land\'{e} factor of 1.17. The distribution of the absolute effective Land\'{e} factors and the line depths is shown in the histograms in Fig. \ref{fig:lines}. We show the absolute value of the Land\'{e} factors because Zeeman splitting in Stokes I does not depend on the sign.

The correlations found in this study are negligibly dependent on the exact choice of the velocity grid width, the model-spectrum deviation, or the minimal and maximal depth of the included lines. Note that, as in the MM-LSD technique, we ensure that the same stellar absorption lines are included for all spectra.

\subsection{Extraction from S2D spectra}

We ran the residual fitting technique described in Section \ref{s:methods} on all echelle order spectra individually and combined the extracted \dabb values by computing the weighted mean of the extracted values of each order. The weight of each order was set to the sum of the inverse squared uncertainties of all included fluxes. The same weights were used in \citet{Lienhard_2022} to combine the common profiles of the individual orders. We also tested running the extraction code on all orders simultaneously. This approach yielded marginally lower correlations with \bobs.

\begin{figure}
    \centering
    \includegraphics[width=\columnwidth]{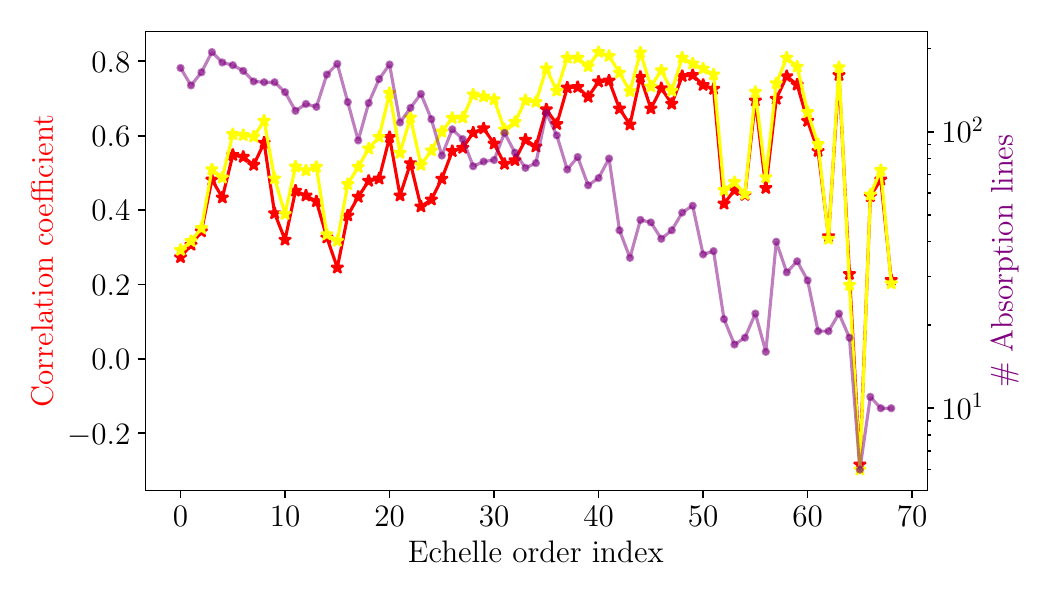}
    \caption{The red (yellow) stars show the Pearson correlation of the order \dabb values with \bobs before (after) correcting for the Sun-Earth viewing angle effect. The purple dots indicate the number of included absorption lines per order.}
    \label{fig:order_correlations}
\end{figure}

Overall we find a good correlation of the extracted \dabb with \bobs from SDO for each order except order 65 (around 6650 \AA) where there are only very few included absorption lines. The Pearson correlation coefficients of \dabb with \bobs for each order are displayed in Fig. \ref{fig:order_correlations}. The central orders between 5000 and 5500 \AA, where we find a high number of isolated absorption lines at high SNR, correlate best with \bobs and also get the highest statistical weight. For lower orders, the absorption lines are more often blended, affected by noise, and the continuum is expected to be less precise. At wavelengths beyond 5500 \AA\ the number of included absorption lines can reach low values leading to high scatter in the extracted order \dabb. Since these orders have a very low statistical weight, their impact on the final \dabb is minor.

\begin{figure}
    \centering
    \includegraphics[width=\columnwidth]{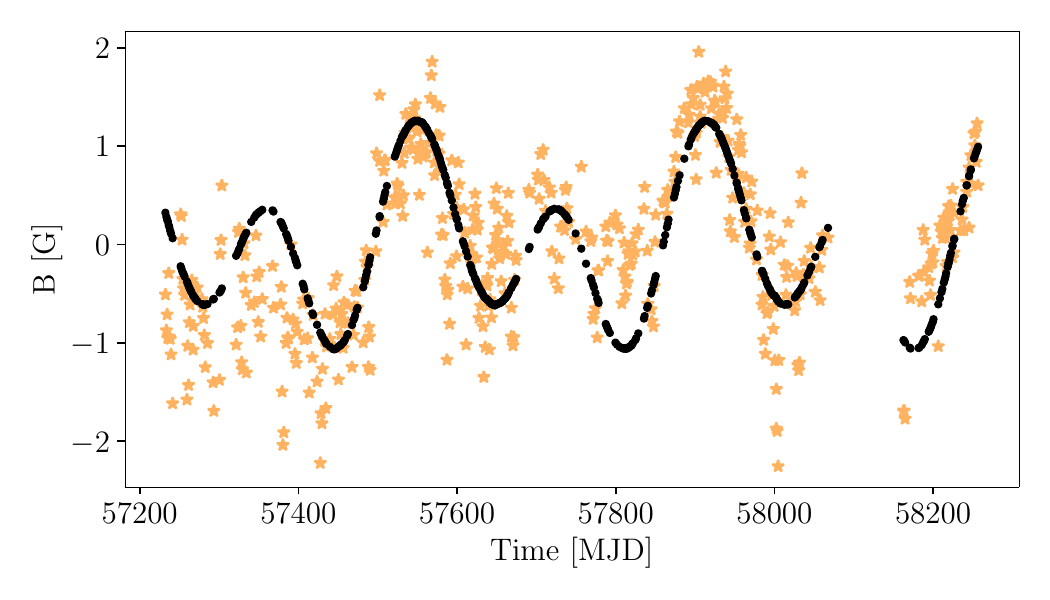}
    \caption{Difference between the scaled \dabb and \bobs (orange stars) and the model for the HARPS-N line width variations (black dots) caused by the varying viewing angle on the Sun.}
    \label{fig:widthcorrection}
\end{figure}

We found it necessary to correct for an observational effect unique to the Sun. As the Earth revolves around the Sun on a slightly eccentric orbit that is inclined relative to the Sun's axis of rotation, the observed rotational line broadening evolves over the course of a year. The width variation is present in the HARPS-N solar spectra and can be modelled as a double-sinusoid \citep{Collier_Cameron_2019}. This signal results in a strong 182-day peak in the periodogram of the width measurements. Our method, being susceptible to line width variations, is susceptible to this viewing angle effect too. However, spectra of other stars are not affected and therefore we remove this signal to get a more realistic estimate of the expected proxy performance for other stars. To remove the double-sinusoid signal, we find the scaling factor that eliminates the 182-day signal from the time series when we subtract the expected line width variation times this scaling factor from the \dabb time series. This is achieved by minimising the power of the respective peak in the Bayesian generalised Lomb-Scargle periodogram \citep{Mortier_2015}. We show the model and the difference between \dabb and \bobs in Fig. \ref{fig:widthcorrection}. Note that we do not use \bobs to compute this correction factor, but the additional signal is most obvious once the magnetic signal has been removed. The correlation coefficients of the individual orders increase by on average 0.06 after applying this correction.
The overall correlation of \dabb with \bobs increases from 0.8 to 0.914 after the removal of this spurious signal. The CCF contrast and FWHM that we use in Section \ref{s:results} have also been corrected for the viewing angle effect following \citet{Collier_Cameron_2019}.

\subsection{Injection-recovery test}

We generated mock spectra to validate our approach given our assumptions. Each of these mock spectra is based on our real spectra. This means that the wavelength solution, the uncertainties and the SNR of the $\text{i}^{\text{th}}$ real spectrum correspond to those of the $\text{i}^{\text{th}}$ mock spectrum.
To generate the mock fluxes, we produced a new line list by splitting each VALD3 line into a $\pi$ and two $\sigma$ components and convolved this list with a normalised Gaussian profile scaled by the depth of the respective component as in Eq. \ref{eq:mag}. We set the field strength in the active regions to 1.5 kG and assumed a factor $\alpha$ between 0.005 and 0.015. We made the same assumptions as in Section \ref{s:methods}. This means that we assume a homogeneous distribution of active regions making the average line-of-sight angle constant and absorbing this factor into $\alpha$. Given the angle-dependent factor in Eq. \ref{eq:seares1} and assuming a characteristic angle $\langle\theta\rangle$ of 34$^{\circ}$ as in \citet{Marcy_1982}, the values of $\alpha$ chosen here correspond to about a filling factor of active regions of about 2 per cent which is within the plage filling factors presented in \citet{Milbourne_2021}.

\begin{figure}
    \centering
    \includegraphics[width=\columnwidth]{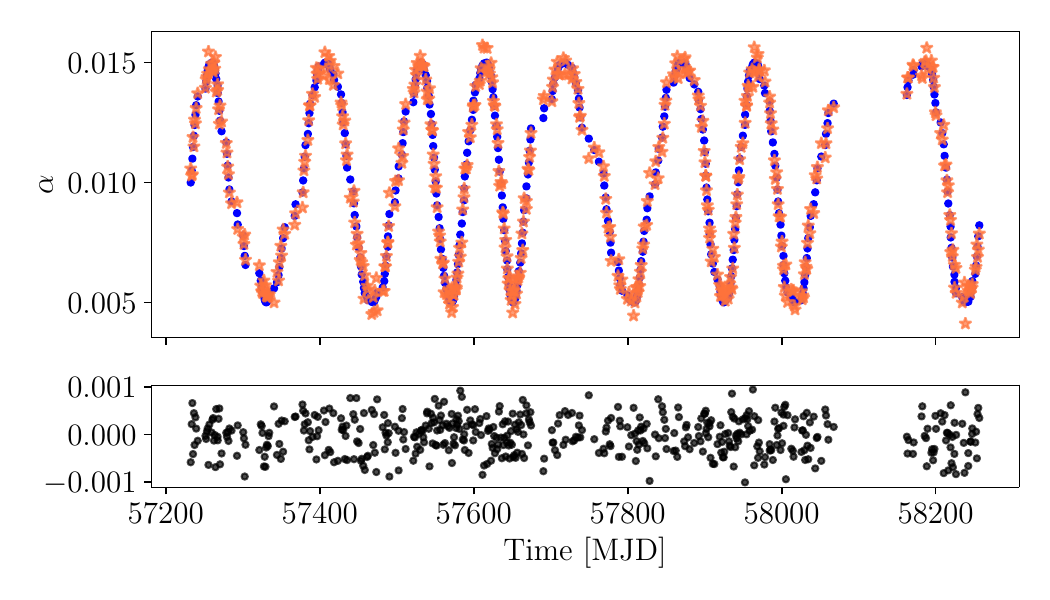}
    \caption{\textit{Upper panel}: Scaled extracted \dabb (orange stars) and comparison to injected values of $\alpha$ (blue dots) assuming a two-component model with the magnetic field strength in the active region being 1.5 kG. \textit{Lower panel}: Difference between scaled \dabb and injected values.}
    \label{fig:simulation_alpha}
\end{figure}

We added Gaussian noise, with the standard deviation of each flux equal to its associated uncertainty estimate, to the spectra. As shown in Fig. \ref{fig:simulation_alpha}, the LSD extraction of \dabb based on the Taylor-expanded Hermite-Gaussian expression successfully retrieves the injected values with no systematic differences and minimal scatter.

\subsection{RV extraction}
It is tempting to simultaneously model the RV and the magnetic effect within the residual profiles to obtain an RV estimate that is less affected by (1) stellar RV effects and (2) effects specific to the RV extraction.\\
The RV impact of stellar activity is expected to be reduced if we allow the spectral component emerging from the active regions to vary in relative strength, to be Zeeman broadened, and to be shifted in RV.
RV extraction effects, on the other hand, are conceivable to originate from the Zeeman-induced width variations of individual lines leading to a varying degree of line blending, for example. This can cause shape variations of the LSD common profile or the CCF because neither of these methods perfectly models line blends. Zeeman intensification can furthermore influence the weight of magnetically sensitive absorption lines in the computation of the common profile. This makes the contribution of these lines to the common profile dependent on the magnetic field and thus produces an activity-dependent contribution to the extracted RV. Correcting for these RV extraction effects cannot remove the magnetically-driven convective RV signals or RV variations caused by the inhomogeneous brightness distribution on the stellar surface. Equally, modelling the photometric effect requires an additional step even if we allow the emission from active regions to vary in strength and Doppler shift within our model.

The difference between two absorption lines that are Doppler shifted relative to each other is well described by the first Hermite-Gaussian polynomial and scales linearly with the RV, as shown in \citet{Holzer_2020}. Using the weak-field triplet splitting model, it can be shown that the RV extracted using the first Hermite-Gaussian polynomial is largely unaffected by Zeeman splitting. This may partially explain the lower RV scatter obtained in \citet{Holzer_2020}. With our line selection including partially blended lines and imperfect line depth estimates, we were able to extract RVs using this method. However, the RV semi-amplitudes were reduced which prohibited meaningful scatter analyses without discarding more data and degrading \dabb. 
The reason for the suppressed RV amplitude can be deduced directly from Eq. 18 in \citet{Holzer_2020}. This expression assumes line additivity which is problematic for the partially blended lines which we included in our analysis.

\section{Discussion} \label{s:results}
In this Section, we present the magnetic flux proxy \dabb as extracted from the HARPS-N solar spectra and compare it with other available activity indicators and RVs.

\subsection{Comparisons of activity indicators}
\begin{figure}
    \centering
    \includegraphics[width=\columnwidth]{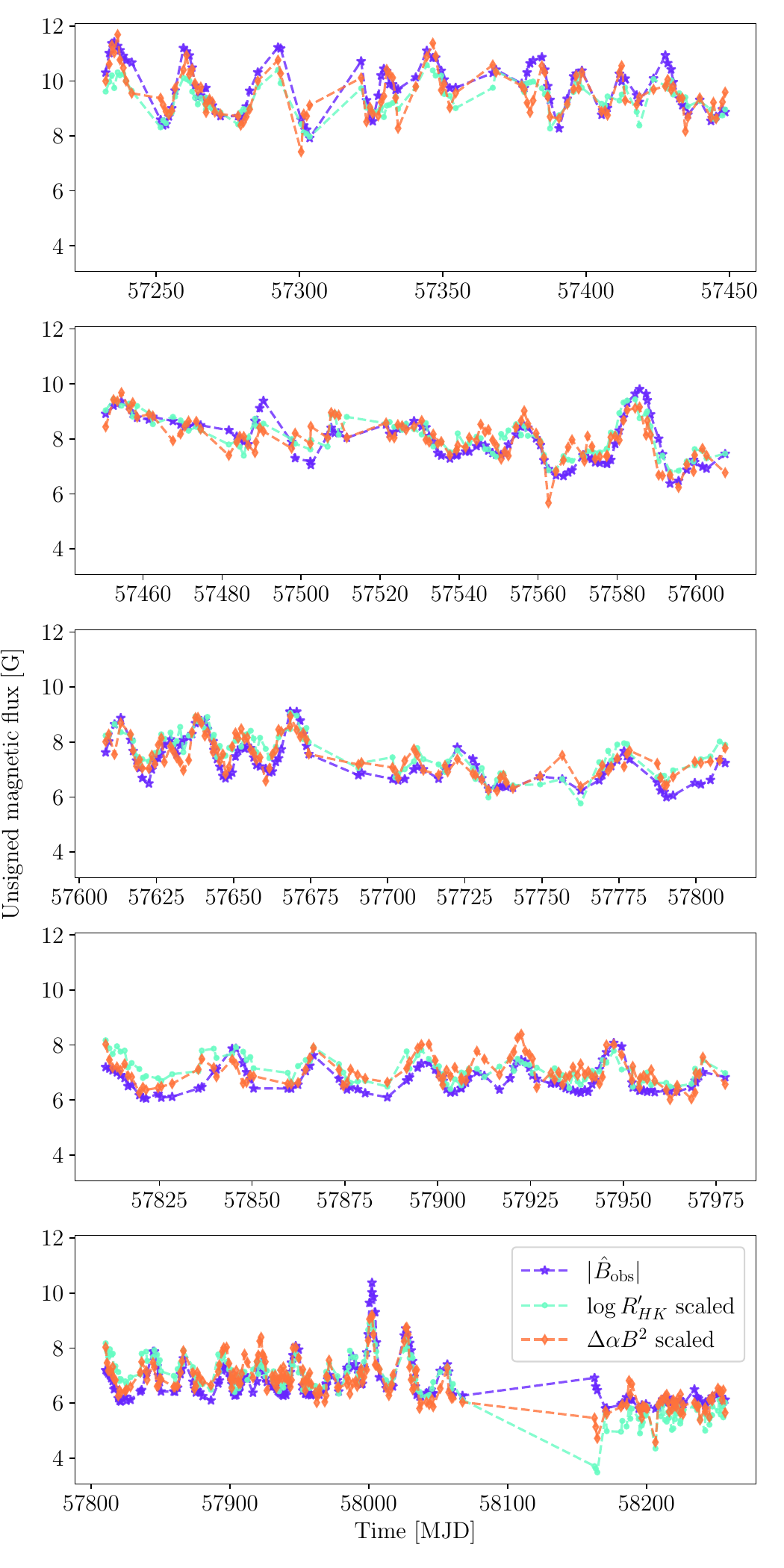}
    \caption{Evolution of \dabb and \rhk, extracted from the HARPS-N solar spectra, and \bobs from HMI/SDO from July 2015 to May 2018.}
    \label{fig:comparison}
\end{figure}

In Fig. \ref{fig:comparison}, we show the complete time series of \dabb and \rhk extracted from the HARPS-N S2D spectra and \bobs from SDO. The overall evolution of the magnetic flux and the quasi-periodic variations on the solar rotational timescale are well traced by all three indicators.

We further evaluated how well our new indicator, \dabb, traces the HARPS-N RV variations extracted using the CCF technique and compare it to the SDO \bobs and other standard activity indicators (\rhk, BIS, FWHM, contrast). For this, we split the data into chunks of a given duration and computed the Pearson correlation coefficient between the RVs and the indicators within these chunks.

\begin{figure}
    \centering
    \includegraphics[width=\columnwidth]{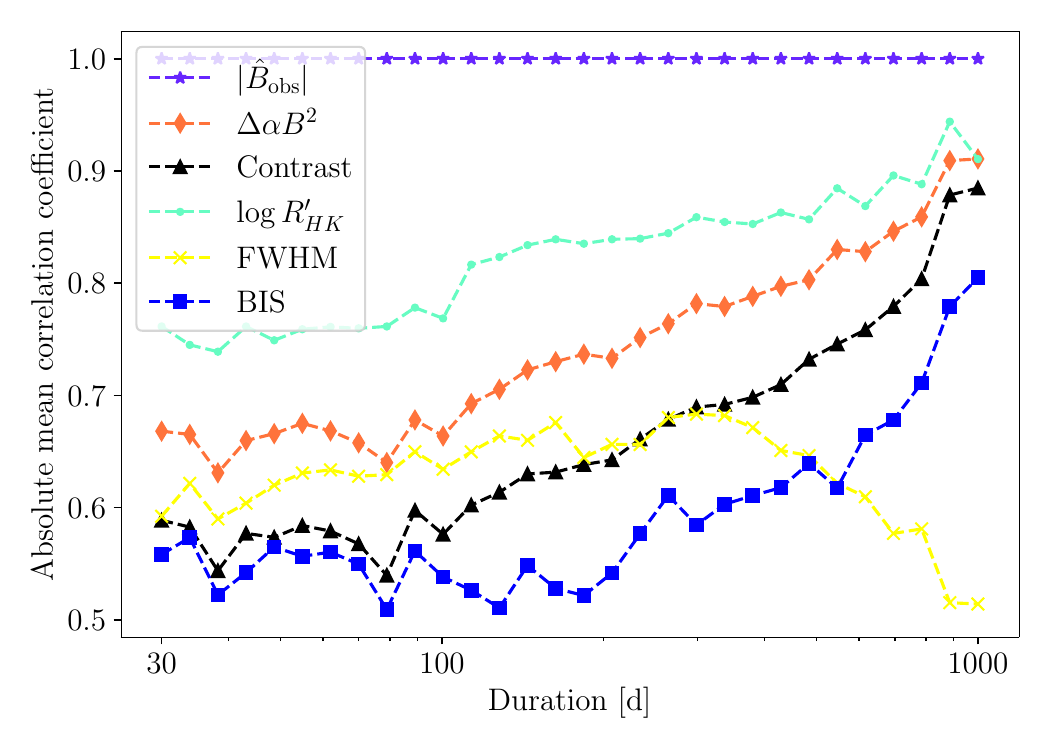}
    \caption{Absolute mean Pearson correlation coefficient of the five activity indices derived from HARPS-N spectra with \bobs for data chunks covering a fixed time span.}
    \label{fig:bcorrs}
\end{figure}

To divide the data with timestamps $t_1$, $t_2$ ... $t_n$ into intervals of $d$ days, we proceeded as follows. As the first chunk of data, we select all measurements taken between $t_1$ and $t_1 +d$. We then computed the Pearson correlation coefficient if this interval contained at least $\frac{d}{3}$ measurements and if these covered at least 90 per cent of the duration $d$. Next, we selected all measurements between $t_1+d/3$ and $t_1 + d/3 +d$, proceeding the same way as above. We thus get a list of Pearson correlation coefficients for data chunks of duration $d$.
In the following Figures \ref{fig:rvcorrs}, \ref{fig:rvcorrs_lsd}, and \ref{fig:bcorrs}, we display the absolute value of the mean of the Pearson correlation coefficients for each time scale $d$.
We included all chunks of data irrespective of whether stellar activity impacted the data significantly. Excluding chunks with RV RMS below 1 \ms did not significantly alter our results.
The correlation analyses were also computed including only chunks of data with RV RMS greater than 1.5 \ms. For these chunks of data, stellar activity is expected to be the dominant contributor to the RV variability with the photon noise contribution being around 0.2 \ms and the long-term instrumental stability of HARPS-N being near 1 \ms. The respective results are shown in Appendix \ref{app:correlations_150cm}.

In Fig. \ref{fig:rvcorrs}, we show the absolute value of the mean of the correlation coefficients computed for these chunks of fixed duration. We computed the absolute value because the contrast is anti-correlated with \bobs. This has also been noted in \citet{Costes_2021}.
The lower correlation coefficients towards shorter timescales can partially be explained by phase offsets between the RV and the indicators, mentioned e.g. in \citet{Collier_Cameron_2019}. These offsets are believed to originate from geometric effects and are therefore expected to be a function of the stellar rotation period. Higher correlations can be achieved by accounting for the time delay between the RVs and the activity indicators.

Unsurprisingly, \bobs traces the RV variations best as it originates from a dedicated magnetic field measurement with extremely high SNR resulting from combining the information from all relevant pixels, and is not affected by the Earth's atmosphere.
\dabb shows a consistently higher correlation with the RV variations as compared to the classical activity indicators.
The contrast shows high correlations with the RVs too, but lower correlations with \bobs as compared to \dabb. It is not surprising that the contrast and \dabb show some similarities because Zeeman broadening alters the depth of absorption lines where the signal-to-noise is highest.
Altogether, his makes \dabb the best known activity indicator that can be extracted from Stokes I spectra simultaneously with the RV for solar-type stars.

One particular advantage of \dabb is that it can easily be extended to include other absorption lines and is not restricted to particular wavelength ranges of detectors. \rhk, in contrast, depends on collecting enough photons where the line reversal emerges around 4000 \AA. 
Furthermore, \rhk weighs the contributions of surface areas differently to the RV extraction because the used wavelength regime, and therefore the limb-darkening, differ. \dabb, on the other hand, weighs the individual echelle orders equivalently to the RV extraction. 
Another difference between \dabb and \rhk is that the flux reversal in the Ca II H and K lines forms above the photosphere, whereas the Zeeman splitting signature traces the magnetic fields in the photosphere where the absorption lines form. This also leads to \rhk essentially seeing a different field distribution since the magnetic field expands with height \citep{Wiegelmann_2014}.
Lastly, \rhk shows a non-linear correlation with the magnetic flux \citep[e.g.][]{Schrijver_1989,Chatzistergos_2019}, whereas we expect \dabb to scale linearly within our model assumptions.

The same comparison for the MM-LSD RVs in Fig. \ref{fig:rvcorrs_lsd} shows generally lower correlation coefficients. Strikingly, the BIS is significantly less correlated with the MM-LSD RVs than with the CCF RVs. This can be partially due to the BIS being calculated from the CCFs themselves rather than the MM-LSD common profiles. However, this would indicate that the BIS is method-specific and therefore less useful as a general diagnostic for the stellar surface conditions.
Alternatively, this may indicate that the line selection for the CCF method is more susceptible to bisector variations.
We performed a simple linear detrending of the RVs with \dabb to estimate the minimal RV RMS improvement as linear detrending is a crude method \citep[for an overview on stellar activity mitigation, see][]{Zhao_2022}. The RMS of the CCF RVs decreased from 2.01 \ms to 1.25 \ms. The RMS of the MM-LSD RVs reduced from 1.65 \ms to 1.22 \ms. The very similar final RMS values indicate that MM-LSD already picks up less stellar activity as compared to the CCF method. This is in agreement with evidence presented in \citet{Lienhard_2022}.

In Fig. \ref{fig:bcorrs}, the correlation of the classical activity indicators and \dabb with \bobs is displayed. \rhk traces the \bobs variations better than \dabb. Both indicators show a high correlation especially for timescales longer than 200 days and trace the variations of \bobs better than the other classical activity indicators. This indicates that \rhk, despite originating in principle in the chromosphere, and \dabb are good indicators for photospheric magnetic flux variations.

\begin{figure}
    \centering
    \includegraphics[width=\columnwidth]{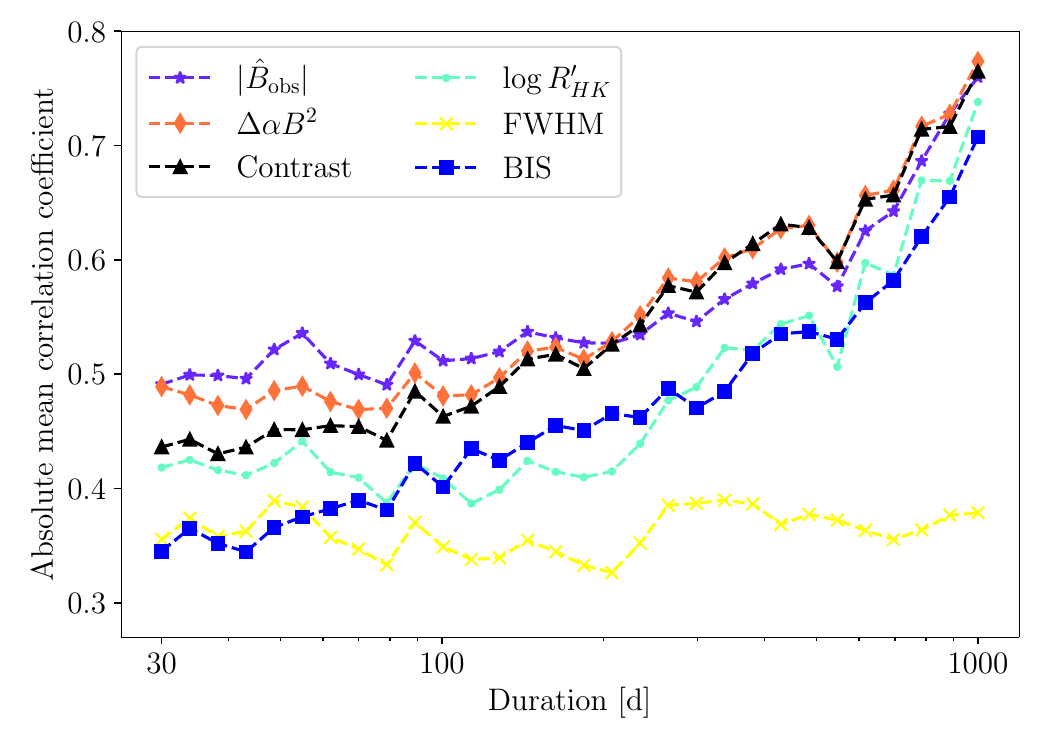}
    \caption{Absolute value of the mean Pearson correlation coefficient of the six activity indices with the heliocentric CCF RVs for data chunks covering a fixed time span.}
    \label{fig:rvcorrs}
\end{figure}

\begin{figure}
    \centering
    \includegraphics[width=\columnwidth]{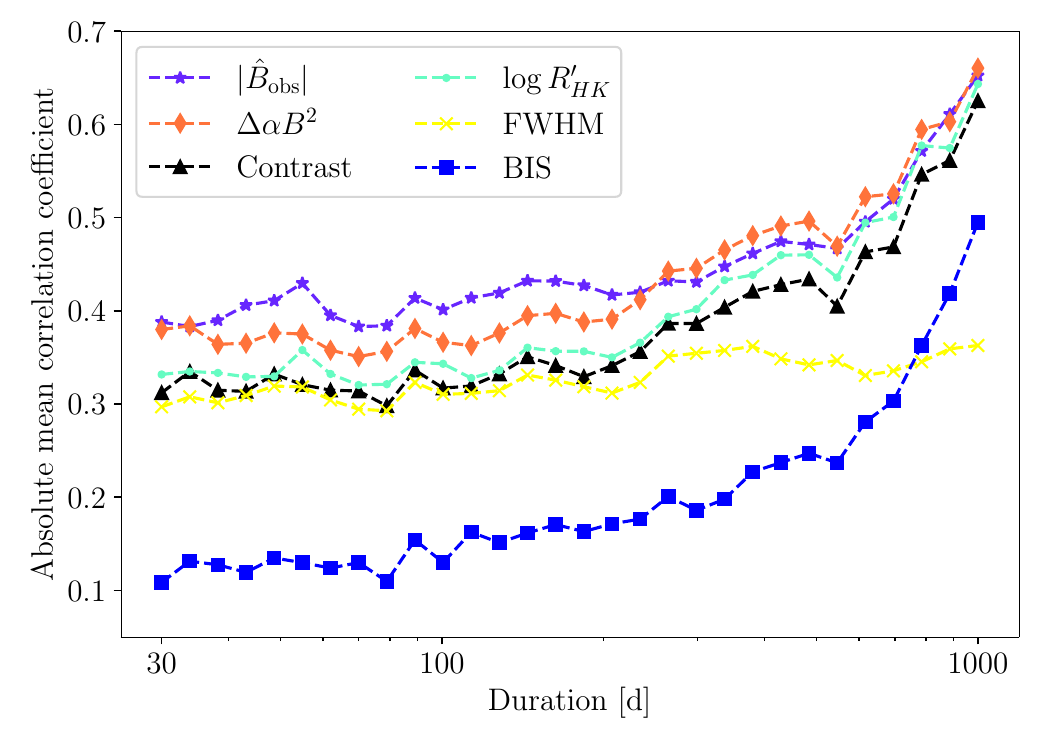}
    \caption{Absolute value of the mean Pearson correlation coefficient of the six activity indices with the heliocentric MM-LSD RVs for data chunks covering a fixed time span.}
    \label{fig:rvcorrs_lsd}
\end{figure}

\subsection{Signal-to-noise ratio dependence} \label{ss:snr_dependence}

The extracted \dabb does not depend crucially on the SNR of the spectra. Neglecting all but photon noise, we can add normally distributed noise to the spectra to simulate spectra at lower SNR. For each flux value, we randomly sampled from a Gaussian distribution with a mean of 0 and the variance equal to the squared uncertainty multiplied by a factor. To double the noise and simulate spectra with half of the original SNR, we set the mentioned factor to 1. The uncertainty estimates were adjusted accordingly.

By doubling the noise, we get a median SNR of 190, which is at the lower end of the expected SNR for very bright targets with visual magnitude of about 6. For these spectra, we still get a Pearson correlation coefficient of 0.91 for \dabb with \bobs overall. This is virtually identical to the Pearson correlation coefficient of 0.914 that we computed for the original spectra with an SNR of about 360.
On the rotational timescale of 30 days, the mean correlation coefficient with \bobs reduces negligibly from 0.67 to 0.66 as we double the noise, whereas the mean correlation coefficient with the CCF RVs reduces from 0.49 to 0.48.

For simulated spectra with median SNR equal to 60, we get a very good overall correlation with \bobs of 0.89. The mean correlation with the \bobs on the rotational timescale is slightly reduced from 0.67 to 0.56. The mean correlation with the CCF RVs reduces from 0.49 to 0.42.
It follows that the described approach is negligibly affected by white noise within the typical SNR range achieved for RV studies.

\section{Conclusions} \label{s:conclusion}

We showed that the difference between a Zeeman-split and an unsplit absorption line is well modelled by Hermite Gaussian polynomials with Taylor-expanded coefficients. The expansion provided a residual model scaling with line-specific factors that allowed us to use the LSD framework to condense the information contained in over 4000 absorption lines and extract \dabb. This indicator represents the hemispherically averaged unsigned magnetic flux variations. We found a correlation of \dabb with \bobs from SDO of 0.91 overall and about 0.67 on the rotational timescale. Importantly, we find a minimal dependence of \dabb on the SNR of the spectra, which makes it a prime activity index even for faint stars. Also, we show in Section \ref{s:results} that \dabb correlates better with activity-induced RV variations than the classical activity indicators, on the rotational timescale and on the overall timescale of 3 years. For solar-type stars, \dabb is thus the best activity tracer known to date that can be simultaneously extracted with the RVs.

We expect this method to perform better if longer wavelength regions are included, provided that the telluric lines are properly corrected and there is a sufficient number of absorption lines. This is due to the Zeeman effect being more pronounced at longer wavelengths. The proxy for \bobs presented in this study provides many potential avenues for extensions and additional applications. For instance, it can provide a magnetic field estimator for stars for which the Calcium H and K lines are too weak, contaminated, or affected by instrumental effects. Also, this provides a magnetic flux estimator for instruments that do not cover the wavelength range required for extracting \rhk. Furthermore, it provides an independent estimate of the evolution of the magnetic flux. The computation of this indicator can be done on readily available spectra that were recorded for RV purposes. No additional equipment or dedicated observation strategies are required.

Several potential modifications of the present approach are conceivable. For instance, additional line-broadening signals could be modelled simultaneously if temperature and magnetic relations are known.
Regarding the Zeeman signal, the Doppler shift of the various components could be modelled as the magnetic and velocity fields are not independent \citep[e.g.][]{Cegla_2013}. Active regions can thus be Doppler shifted relative to the quiet photospheric regions. On the technical side, instead of comparing the stellar spectra to an averaged stellar spectrum of the same star, simulated spectra could be used.

Lastly, there are many applications and extensions for the presented indicator. The magnetic flux at different heights within the photosphere may be probed by making line selections.
For our next steps, we intend to generalise this method to other stars and assess the performance of the indicator for stellar activity modelling.

\section*{Acknowledgements}
The authors would like to thank the referee Steven Saar for his detailed and insightful comments.
FL gratefully acknowledges a scholarship from the Fondation Zd\u{e}nek et Michaela Bakala. FL would like to thank Prof. J.O. Stenflo for useful discussions. AM acknowledges support from the senior Kavli Institute Fellowships. ACC acknowledges support from STFC consolidated grant numbers ST/R000824/1 and ST/V000861/1, 
and UKSA grant number ST/R003203/1.
BK acknowledges funding from the Euro-
pean Research Council under the European Union’s Horizon 2020
research and innovation programme (grant agreement No 865624,
GPRV).

\section*{Data Availability}

The solar spectra are available on the Data \& Analysis Center for Exoplanets (DACE) web-platform operated by the University of Geneva (\url{https://dace.unige.ch/sun})



\bibliographystyle{mnras}
\bibliography{example} 

\begin{thebibliography}{}
\makeatletter
\relax
\def\mn@urlcharsother{\let\do\@makeother \do\$\do\&\do\#\do\^\do\_\do\%\do\~}
\def\mn@doi{\begingroup\mn@urlcharsother \@ifnextchar [ {\mn@doi@}
  {\mn@doi@[]}}
\def\mn@doi@[#1]#2{\def\@tempa{#1}\ifx\@tempa\@empty \href
  {http://dx.doi.org/#2} {doi:#2}\else \href {http://dx.doi.org/#2} {#1}\fi
  \endgroup}
\def\mn@eprint#1#2{\mn@eprint@#1:#2::\@nil}
\def\mn@eprint@arXiv#1{\href {http://arxiv.org/abs/#1} {{\tt arXiv:#1}}}
\def\mn@eprint@dblp#1{\href {http://dblp.uni-trier.de/rec/bibtex/#1.xml}
  {dblp:#1}}
\def\mn@eprint@#1:#2:#3:#4\@nil{\def\@tempa {#1}\def\@tempb {#2}\def\@tempc
  {#3}\ifx \@tempc \@empty \let \@tempc \@tempb \let \@tempb \@tempa \fi \ifx
  \@tempb \@empty \def\@tempb {arXiv}\fi \@ifundefined
  {mn@eprint@\@tempb}{\@tempb:\@tempc}{\expandafter \expandafter \csname
  mn@eprint@\@tempb\endcsname \expandafter{\@tempc}}}

\bibitem[\protect\citeauthoryear{{Ayres}}{{Ayres}}{2014}]{Ayres_2014}
{Ayres} T.~R.,  2014, \mn@doi [\aj] {10.1088/0004-6256/147/3/59}, \href
  {https://ui.adsabs.harvard.edu/abs/2014AJ....147...59A} {147, 59}

\bibitem[\protect\citeauthoryear{{Baranne} et~al.,}{{Baranne}
  et~al.}{1996}]{Baranne_1996}
{Baranne} A.,  et~al., 1996, \aaps, \href
  {https://ui.adsabs.harvard.edu/abs/1996A&AS..119..373B} {119, 373}

\bibitem[\protect\citeauthoryear{{Basri} \& {Marcy}}{{Basri} \&
  {Marcy}}{1994}]{Basri_1994}
{Basri} G.,  {Marcy} G.~W.,  1994, \mn@doi [\apj] {10.1086/174535}, \href
  {https://ui.adsabs.harvard.edu/abs/1994ApJ...431..844B} {431, 844}

\bibitem[\protect\citeauthoryear{{Bellot Rubio} \& {Orozco Su{\'a}rez}}{{Bellot
  Rubio} \& {Orozco Su{\'a}rez}}{2019}]{Bellot_Rubio_2019}
{Bellot Rubio} L.,  {Orozco Su{\'a}rez} D.,  2019, \mn@doi [Living Reviews in
  Solar Physics] {10.1007/s41116-018-0017-1}, \href
  {https://ui.adsabs.harvard.edu/abs/2019LRSP...16....1B} {16, 1}

\bibitem[\protect\citeauthoryear{{Bertaux}, {Lallement}, {Ferron}  \&
  {Boonne}}{{Bertaux} et~al.}{2014}]{Bertaux_2014}
{Bertaux} J.-L.,  {Lallement} R.,  {Ferron} S.,   {Boonne} C.,  2014, in 13th
  International HITRAN Conference. p.~8, \mn@doi{10.5281/zenodo.11110}

\bibitem[\protect\citeauthoryear{{Brault} \& {Noyes}}{{Brault} \&
  {Noyes}}{1983}]{Brault_1983}
{Brault} J.,  {Noyes} R.,  1983, \mn@doi [\apjl] {10.1086/184056}, \href
  {https://ui.adsabs.harvard.edu/abs/1983ApJ...269L..61B} {269, L61}

\bibitem[\protect\citeauthoryear{{Bruls} \& {Solanki}}{{Bruls} \&
  {Solanki}}{1995}]{Bruls_1995}
{Bruls} J.~H.~M.~J.,  {Solanki} S.~K.,  1995, \aap, \href
  {https://ui.adsabs.harvard.edu/abs/1995A&A...293..240B} {293, 240}

\bibitem[\protect\citeauthoryear{{Buehler}, {Lagg}, {Solanki}  \& {van
  Noort}}{{Buehler} et~al.}{2015}]{Buehler_2015}
{Buehler} D.,  {Lagg} A.,  {Solanki} S.,   {van Noort} M.,  2015, \mn@doi
  [\aap] {10.1051/0004-6361/201424970}, \href
  {https://ui.adsabs.harvard.edu/abs/2015A&A...576A..27B} {576, A27}

\bibitem[\protect\citeauthoryear{{Buehler}, {Lagg}, {van Noort}  \&
  {Solanki}}{{Buehler} et~al.}{2019}]{Buehler_2019}
{Buehler} D.,  {Lagg} A.,  {van Noort} M.,   {Solanki} S.~K.,  2019, \mn@doi
  [\aap] {10.1051/0004-6361/201833585}, \href
  {https://ui.adsabs.harvard.edu/abs/2019A&A...630A..86B} {630, A86}

\bibitem[\protect\citeauthoryear{{Cegla}, {Shelyag}, {Watson}  \&
  {Mathioudakis}}{{Cegla} et~al.}{2013}]{Cegla_2013}
{Cegla} H.~M.,  {Shelyag} S.,  {Watson} C.~A.,   {Mathioudakis} M.,  2013,
  \mn@doi [\apj] {10.1088/0004-637X/763/2/95}, \href
  {https://ui.adsabs.harvard.edu/abs/2013ApJ...763...95C} {763, 95}

\bibitem[\protect\citeauthoryear{{Cegla}, {Watson}, {Shelyag}, {Mathioudakis}
  \& {Moutari}}{{Cegla} et~al.}{2019}]{Cegla_2019}
{Cegla} H.~M.,  {Watson} C.~A.,  {Shelyag} S.,  {Mathioudakis} M.,   {Moutari}
  S.,  2019, \mn@doi [\apj] {10.3847/1538-4357/ab16d3}, \href
  {https://ui.adsabs.harvard.edu/abs/2019ApJ...879...55C} {879, 55}

\bibitem[\protect\citeauthoryear{{Chaplin}, {Cegla}, {Watson}, {Davies}  \&
  {Ball}}{{Chaplin} et~al.}{2019}]{Chaplin_2019}
{Chaplin} W.~J.,  {Cegla} H.~M.,  {Watson} C.~A.,  {Davies} G.~R.,   {Ball}
  W.~H.,  2019, \mn@doi [\aj] {10.3847/1538-3881/ab0c01}, \href
  {https://ui.adsabs.harvard.edu/abs/2019AJ....157..163C} {157, 163}

\bibitem[\protect\citeauthoryear{{Chatzistergos}, {Ermolli}, {Solanki},
  {Krivova}, {Giorgi}  \& {Yeo}}{{Chatzistergos}
  et~al.}{2019}]{Chatzistergos_2019}
{Chatzistergos} T.,  {Ermolli} I.,  {Solanki} S.~K.,  {Krivova} N.~A.,
  {Giorgi} F.,   {Yeo} K.~L.,  2019, \mn@doi [\aap]
  {10.1051/0004-6361/201935131}, \href
  {https://ui.adsabs.harvard.edu/abs/2019A&A...626A.114C} {626, A114}

\bibitem[\protect\citeauthoryear{{Collier Cameron} et~al.,}{{Collier Cameron}
  et~al.}{2019}]{Collier_Cameron_2019}
{Collier Cameron} A.,  et~al., 2019, \mn@doi [\mnras] {10.1093/mnras/stz1215},
  \href {https://ui.adsabs.harvard.edu/abs/2019MNRAS.487.1082C} {487, 1082}

\bibitem[\protect\citeauthoryear{{Cosentino} et~al.,}{{Cosentino}
  et~al.}{2012}]{Cosentino_2012}
{Cosentino} R.,  et~al., 2012, in {McLean} I.~S.,  {Ramsay} S.~K.,   {Takami}
  H.,  eds,  Society of Photo-Optical Instrumentation Engineers (SPIE)
  Conference Series Vol. 8446, Ground-based and Airborne Instrumentation for
  Astronomy IV. p. 84461V, \mn@doi{10.1117/12.925738}

\bibitem[\protect\citeauthoryear{{Cosentino} et~al.,}{{Cosentino}
  et~al.}{2014}]{Cosentino_2014}
{Cosentino} R.,  et~al., 2014, in {Ramsay} S.~K.,  {McLean} I.~S.,   {Takami}
  H.,  eds,  Society of Photo-Optical Instrumentation Engineers (SPIE)
  Conference Series Vol. 9147, Ground-based and Airborne Instrumentation for
  Astronomy V. p. 91478C, \mn@doi{10.1117/12.2055813}

\bibitem[\protect\citeauthoryear{{Costes} et~al.,}{{Costes}
  et~al.}{2021}]{Costes_2021}
{Costes} J.~C.,  et~al., 2021, \mn@doi [\mnras] {10.1093/mnras/stab1183}, \href
  {https://ui.adsabs.harvard.edu/abs/2021MNRAS.505..830C} {505, 830}

\bibitem[\protect\citeauthoryear{{Couvidat} et~al.,}{{Couvidat}
  et~al.}{2016}]{Couvidat_2016}
{Couvidat} S.,  et~al., 2016, \mn@doi [\solphys] {10.1007/s11207-016-0957-3},
  \href {https://ui.adsabs.harvard.edu/abs/2016SoPh..291.1887C} {291, 1887}

\bibitem[\protect\citeauthoryear{{Crass} et~al.,}{{Crass}
  et~al.}{2021}]{Crass_2021}
{Crass} J.,  et~al., 2021, arXiv e-prints, \href
  {https://ui.adsabs.harvard.edu/abs/2021arXiv210714291C} {p. arXiv:2107.14291}

\bibitem[\protect\citeauthoryear{{Cretignier}, {Francfort}, {Dumusque},
  {Allart}  \& {Pepe}}{{Cretignier} et~al.}{2020}]{Cretignier_2020}
{Cretignier} M.,  {Francfort} J.,  {Dumusque} X.,  {Allart} R.,   {Pepe} F.,
  2020, \mn@doi [\aap] {10.1051/0004-6361/202037722}, \href
  {https://ui.adsabs.harvard.edu/abs/2020A&A...640A..42C} {640, A42}

\bibitem[\protect\citeauthoryear{{Cretignier}, {Dumusque}, {Hara}  \&
  {Pepe}}{{Cretignier} et~al.}{2021}]{Cretignier_2021}
{Cretignier} M.,  {Dumusque} X.,  {Hara} N.~C.,   {Pepe} F.,  2021, \mn@doi
  [\aap] {10.1051/0004-6361/202140986}, \href
  {https://ui.adsabs.harvard.edu/abs/2021A&A...653A..43C} {653, A43}

\bibitem[\protect\citeauthoryear{{Cunha}, {Santos}, {Figueira}, {Santerne},
  {Bertaux}  \& {Lovis}}{{Cunha} et~al.}{2014}]{Cunha_2014}
{Cunha} D.,  {Santos} N.~C.,  {Figueira} P.,  {Santerne} A.,  {Bertaux} J.~L.,
   {Lovis} C.,  2014, \mn@doi [\aap] {10.1051/0004-6361/201423723}, \href
  {https://ui.adsabs.harvard.edu/abs/2014A&A...568A..35C} {568, A35}

\bibitem[\protect\citeauthoryear{{Desort}, {Lagrange}, {Galland}, {Udry}  \&
  {Mayor}}{{Desort} et~al.}{2007}]{Desort_2007}
{Desort} M.,  {Lagrange} A.~M.,  {Galland} F.,  {Udry} S.,   {Mayor} M.,  2007,
  \mn@doi [\aap] {10.1051/0004-6361:20078144}, \href
  {https://ui.adsabs.harvard.edu/abs/2007A&A...473..983D} {473, 983}

\bibitem[\protect\citeauthoryear{{Donati}, {Semel}  \& {Praderie}}{{Donati}
  et~al.}{1989}]{Donati_zdi_1989}
{Donati} J.~F.,  {Semel} M.,   {Praderie} F.,  1989, \aap, \href
  {https://ui.adsabs.harvard.edu/abs/1989A&A...225..467D} {225, 467}

\bibitem[\protect\citeauthoryear{{Donati}, {Semel}, {Carter}, {Rees}  \&
  {Collier Cameron}}{{Donati} et~al.}{1997}]{Donati_1997}
{Donati} J.~F.,  {Semel} M.,  {Carter} B.~D.,  {Rees} D.~E.,   {Collier
  Cameron} A.,  1997, \mn@doi [\mnras] {10.1093/mnras/291.4.658}, \href
  {https://ui.adsabs.harvard.edu/abs/1997MNRAS.291..658D} {291, 658}

\bibitem[\protect\citeauthoryear{{Donati} et~al.,}{{Donati}
  et~al.}{2006}]{Donati_2006}
{Donati} J.~F.,  et~al., 2006, \mn@doi [\mnras]
  {10.1111/j.1365-2966.2006.10558.x}, \href
  {https://ui.adsabs.harvard.edu/abs/2006MNRAS.370..629D} {370, 629}

\bibitem[\protect\citeauthoryear{{Dravins}}{{Dravins}}{1982}]{Dravins_1982}
{Dravins} D.,  1982, \mn@doi [\araa] {10.1146/annurev.aa.20.090182.000425},
  \href {https://ui.adsabs.harvard.edu/abs/1982ARA&A..20...61D} {20, 61}

\bibitem[\protect\citeauthoryear{{Dumusque}, {Udry}, {Lovis}, {Santos}  \&
  {Monteiro}}{{Dumusque} et~al.}{2011}]{Dumusque_2011}
{Dumusque} X.,  {Udry} S.,  {Lovis} C.,  {Santos} N.~C.,   {Monteiro}
  M.~J.~P.~F.~G.,  2011, \mn@doi [\aap] {10.1051/0004-6361/201014097}, \href
  {https://ui.adsabs.harvard.edu/abs/2011A&A...525A.140D} {525, A140}

\bibitem[\protect\citeauthoryear{{Dumusque} et~al.,}{{Dumusque}
  et~al.}{2015}]{Dumusque_2015}
{Dumusque} X.,  et~al., 2015, \mn@doi [\apjl] {10.1088/2041-8205/814/2/L21},
  \href {https://ui.adsabs.harvard.edu/abs/2015ApJ...814L..21D} {814, L21}

\bibitem[\protect\citeauthoryear{{Dumusque} et~al.,}{{Dumusque}
  et~al.}{2021}]{Dumusque_2021}
{Dumusque} X.,  et~al., 2021, \mn@doi [\aap] {10.1051/0004-6361/202039350},
  \href {https://ui.adsabs.harvard.edu/abs/2021A&A...648A.103D} {648, A103}

\bibitem[\protect\citeauthoryear{{Ervin} et~al.,}{{Ervin}
  et~al.}{2022}]{Ervin_2022}
{Ervin} T.,  et~al., 2022, \mn@doi [\aj] {10.3847/1538-3881/ac67e6}, \href
  {https://ui.adsabs.harvard.edu/abs/2022AJ....163..272E} {163, 272}

\bibitem[\protect\citeauthoryear{{Faria} et~al.,}{{Faria}
  et~al.}{2022}]{Faria_2022}
{Faria} J.~P.,  et~al., 2022, \mn@doi [\aap] {10.1051/0004-6361/202142337},
  \href {https://ui.adsabs.harvard.edu/abs/2022A&A...658A.115F} {658, A115}

\bibitem[\protect\citeauthoryear{{Folsom} et~al.,}{{Folsom}
  et~al.}{2018}]{Folsom_2018}
{Folsom} C.~P.,  et~al., 2018, \mn@doi [\mnras] {10.1093/mnras/stx3021}, \href
  {https://ui.adsabs.harvard.edu/abs/2018MNRAS.474.4956F} {474, 4956}

\bibitem[\protect\citeauthoryear{{Giampapa}, {Golub}  \& {Worden}}{{Giampapa}
  et~al.}{1983}]{Giampapa_1983}
{Giampapa} M.~S.,  {Golub} L.,   {Worden} S.~P.,  1983, \mn@doi [\apjl]
  {10.1086/184041}, \href
  {https://ui.adsabs.harvard.edu/abs/1983ApJ...268L.121G} {268, L121}

\bibitem[\protect\citeauthoryear{{Gomes da Silva}, {Santos}, {Bonfils},
  {Delfosse}, {Forveille}  \& {Udry}}{{Gomes da Silva}
  et~al.}{2011}]{GomesdaSilva_2011}
{Gomes da Silva} J.,  {Santos} N.~C.,  {Bonfils} X.,  {Delfosse} X.,
  {Forveille} T.,   {Udry} S.,  2011, \mn@doi [\aap]
  {10.1051/0004-6361/201116971}, \href
  {https://ui.adsabs.harvard.edu/abs/2011A&A...534A..30G} {534, A30}

\bibitem[\protect\citeauthoryear{{Gray}}{{Gray}}{2005}]{Gray_2005}
{Gray} D.~F.,  2005, {The Observation and Analysis of Stellar Photospheres}

\bibitem[\protect\citeauthoryear{{Hale}}{{Hale}}{1908}]{Hale_1908}
{Hale} G.~E.,  1908, \mn@doi [\apj] {10.1086/141602}, \href
  {https://ui.adsabs.harvard.edu/abs/1908ApJ....28..315H} {28, 315}

\bibitem[\protect\citeauthoryear{{Halverson} et~al.,}{{Halverson}
  et~al.}{2016}]{Halverson_2016}
{Halverson} S.,  et~al., 2016, in {Evans} C.~J.,  {Simard} L.,   {Takami} H.,
  eds,  Society of Photo-Optical Instrumentation Engineers (SPIE) Conference
  Series Vol. 9908, Ground-based and Airborne Instrumentation for Astronomy VI.
  p. 99086P (\mn@eprint {arXiv} {1607.05634}), \mn@doi{10.1117/12.2232761}

\bibitem[\protect\citeauthoryear{{Hanslmeier}, {Nesis}  \&
  {Mattig}}{{Hanslmeier} et~al.}{1991}]{Hanslmeier_1991}
{Hanslmeier} A.,  {Nesis} A.,   {Mattig} W.,  1991, \aap, \href
  {https://ui.adsabs.harvard.edu/abs/1991A&A...251..307H} {251, 307}

\bibitem[\protect\citeauthoryear{{Haywood} et~al.,}{{Haywood}
  et~al.}{2016}]{Haywood_2016}
{Haywood} R.~D.,  et~al., 2016, in 19th Cambridge Workshop on Cool Stars,
  Stellar Systems, and the Sun (CS19). Cambridge Workshop on Cool Stars,
  Stellar Systems, and the Sun.
p.~47, \mn@doi{10.5281/zenodo.55693}

\bibitem[\protect\citeauthoryear{{Haywood} et~al.,}{{Haywood}
  et~al.}{2022}]{Haywood_2022}
{Haywood} R.~D.,  et~al., 2022, \mn@doi [\apj] {10.3847/1538-4357/ac7c12},
  \href {https://ui.adsabs.harvard.edu/abs/2022ApJ...935....6H} {935, 6}

\bibitem[\protect\citeauthoryear{Holzer, Cisewski-Kehe, Fischer  \&
  Zhao}{Holzer et~al.}{2021}]{Holzer_2020}
Holzer P.~H.,  Cisewski-Kehe J.,  Fischer D.,   Zhao L.,  2021, \mn@doi [The
  Annals of Applied Statistics] {10.1214/20-AOAS1406}, 15, 527

\bibitem[\protect\citeauthoryear{{Jefferies}, {Lites}  \&
  {Skumanich}}{{Jefferies} et~al.}{1989}]{Jefferies_1989}
{Jefferies} J.,  {Lites} B.~W.,   {Skumanich} A.,  1989, \mn@doi [\apj]
  {10.1086/167762}, \href
  {https://ui.adsabs.harvard.edu/abs/1989ApJ...343..920J} {343, 920}

\bibitem[\protect\citeauthoryear{{Johns-Krull}, {Valenti}  \&
  {Koresko}}{{Johns-Krull} et~al.}{1999}]{Johns-Krull_1999}
{Johns-Krull} C.~M.,  {Valenti} J.~A.,   {Koresko} C.,  1999, \mn@doi [\apj]
  {10.1086/307128}, \href
  {https://ui.adsabs.harvard.edu/abs/1999ApJ...516..900J} {516, 900}

\bibitem[\protect\citeauthoryear{{Jurgenson}, {Fischer}, {McCracken}, {Sawyer},
  {Szymkowiak}, {Davis}, {Muller}  \& {Santoro}}{{Jurgenson}
  et~al.}{2016}]{Jurgenson_2016}
{Jurgenson} C.,  {Fischer} D.,  {McCracken} T.,  {Sawyer} D.,  {Szymkowiak} A.,
   {Davis} A.,  {Muller} G.,   {Santoro} F.,  2016, in {Evans} C.~J.,  {Simard}
  L.,   {Takami} H.,  eds,  Society of Photo-Optical Instrumentation Engineers
  (SPIE) Conference Series Vol. 9908, Ground-based and Airborne Instrumentation
  for Astronomy VI. p. 99086T (\mn@eprint {arXiv} {1606.04413}),
  \mn@doi{10.1117/12.2233002}

\bibitem[\protect\citeauthoryear{{Kochukhov} \& {Wade}}{{Kochukhov} \&
  {Wade}}{2016}]{Kochukhov_2016}
{Kochukhov} O.,  {Wade} G.~A.,  2016, \mn@doi [\aap]
  {10.1051/0004-6361/201527454}, \href
  {https://ui.adsabs.harvard.edu/abs/2016A&A...586A..30K} {586, A30}

\bibitem[\protect\citeauthoryear{{Kochukhov}, {Makaganiuk}  \&
  {Piskunov}}{{Kochukhov} et~al.}{2010}]{Kochukhov_2010}
{Kochukhov} O.,  {Makaganiuk} V.,   {Piskunov} N.,  2010, \mn@doi [\aap]
  {10.1051/0004-6361/201015429}, \href
  {https://ui.adsabs.harvard.edu/abs/2010A&A...524A...5K} {524, A5}

\bibitem[\protect\citeauthoryear{{Kochukhov} et~al.,}{{Kochukhov}
  et~al.}{2011}]{Kochukhov_2011}
{Kochukhov} O.,  et~al., 2011, \mn@doi [\apjl] {10.1088/2041-8205/732/2/L19},
  \href {https://ui.adsabs.harvard.edu/abs/2011ApJ...732L..19K} {732, L19}

\bibitem[\protect\citeauthoryear{{Kochukhov}, {Hackman}, {Lehtinen}  \&
  {Wehrhahn}}{{Kochukhov} et~al.}{2020}]{Kochukhov_2020}
{Kochukhov} O.,  {Hackman} T.,  {Lehtinen} J.~J.,   {Wehrhahn} A.,  2020,
  \mn@doi [\aap] {10.1051/0004-6361/201937185}, \href
  {https://ui.adsabs.harvard.edu/abs/2020A&A...635A.142K} {635, A142}

\bibitem[\protect\citeauthoryear{{Lagrange}, {Desort}  \& {Meunier}}{{Lagrange}
  et~al.}{2010}]{Lagrange_2010}
{Lagrange} A.~M.,  {Desort} M.,   {Meunier} N.,  2010, \mn@doi [\aap]
  {10.1051/0004-6361/200913071}, \href
  {https://ui.adsabs.harvard.edu/abs/2010A&A...512A..38L} {512, A38}

\bibitem[\protect\citeauthoryear{{Landi Degl'Innocenti}}{{Landi
  Degl'Innocenti}}{1982}]{Landi_1982}
{Landi Degl'Innocenti} E.,  1982, \mn@doi [\solphys] {10.1007/BF00156111},
  \href {https://ui.adsabs.harvard.edu/abs/1982SoPh...77..285L} {77, 285}

\bibitem[\protect\citeauthoryear{{Landolfi} \& {Landi
  Degl'Innocenti}}{{Landolfi} \& {Landi Degl'Innocenti}}{1982}]{Landolfi_1982}
{Landolfi} M.,  {Landi Degl'Innocenti} E.,  1982, \mn@doi [\solphys]
  {10.1007/BF00151615}, \href
  {https://ui.adsabs.harvard.edu/abs/1982SoPh...78..355L} {78, 355}

\bibitem[\protect\citeauthoryear{{Lanza}}{{Lanza}}{2010}]{Lanza_2010}
{Lanza} A.~F.,  2010, in {Kosovichev} A.~G.,  {Andrei} A.~H.,   {Rozelot}
  J.-P.,  eds,  Proceedings of the International Astronomical Union Vol. 264,
  Solar and Stellar Variability: Impact on Earth and Planets. pp 120--129
  (\mn@eprint {arXiv} {0909.4660}), \mn@doi{10.1017/S1743921309992523}

\bibitem[\protect\citeauthoryear{{Leet}, {Fischer}  \& {Valenti}}{{Leet}
  et~al.}{2019}]{Leet_2019}
{Leet} C.,  {Fischer} D.~A.,   {Valenti} J.~A.,  2019, \mn@doi [\aj]
  {10.3847/1538-3881/ab0d86}, \href
  {https://ui.adsabs.harvard.edu/abs/2019AJ....157..187L} {157, 187}

\bibitem[\protect\citeauthoryear{{Lehmann}, {K{\"u}nstler}, {Carroll}  \&
  {Strassmeier}}{{Lehmann} et~al.}{2015}]{Lehmann_2015}
{Lehmann} L.~T.,  {K{\"u}nstler} A.,  {Carroll} T.~A.,   {Strassmeier} K.~G.,
  2015, \mn@doi [Astronomische Nachrichten] {10.1002/asna.201412162}, \href
  {https://ui.adsabs.harvard.edu/abs/2015AN....336..258L} {336, 258}

\bibitem[\protect\citeauthoryear{{Lienhard}, {Mortier}, {Buchhave}, {Collier
  Cameron}, {L{\'o}pez-Morales}, {Sozzetti}, {Watson}  \&
  {Cosentino}}{{Lienhard} et~al.}{2022}]{Lienhard_2022}
{Lienhard} F.,  {Mortier} A.,  {Buchhave} L.,  {Collier Cameron} A.,
  {L{\'o}pez-Morales} M.,  {Sozzetti} A.,  {Watson} C.~A.,   {Cosentino} R.,
  2022, \mn@doi [\mnras] {10.1093/mnras/stac1098}, \href
  {https://ui.adsabs.harvard.edu/abs/2022MNRAS.513.5328L} {513, 5328}

\bibitem[\protect\citeauthoryear{{Marcy}}{{Marcy}}{1982}]{Marcy_1982}
{Marcy} G.~W.,  1982, \mn@doi [\pasp] {10.1086/131097}, \href
  {https://ui.adsabs.harvard.edu/abs/1982PASP...94..989M} {94, 989}

\bibitem[\protect\citeauthoryear{{Mart{\'\i}nez Pillet}, {Lites}  \&
  {Skumanich}}{{Mart{\'\i}nez Pillet} et~al.}{1997}]{MartinezPillet_1997}
{Mart{\'\i}nez Pillet} V.,  {Lites} B.~W.,   {Skumanich} A.,  1997, \mn@doi
  [\apj] {10.1086/303478}, \href
  {https://ui.adsabs.harvard.edu/abs/1997ApJ...474..810M} {474, 810}

\bibitem[\protect\citeauthoryear{{Mayor} et~al.,}{{Mayor}
  et~al.}{2003}]{Mayor_2003}
{Mayor} M.,  et~al., 2003, The Messenger, \href
  {https://ui.adsabs.harvard.edu/abs/2003Msngr.114...20M} {114, 20}

\bibitem[\protect\citeauthoryear{{Medina}, {Johnson}, {Eastman}  \&
  {Cargile}}{{Medina} et~al.}{2018}]{Medina_2018}
{Medina} A.~A.,  {Johnson} J.~A.,  {Eastman} J.~D.,   {Cargile} P.~A.,  2018,
  \mn@doi [\apj] {10.3847/1538-4357/aadf82}, \href
  {https://ui.adsabs.harvard.edu/abs/2018ApJ...867...32M} {867, 32}

\bibitem[\protect\citeauthoryear{{Meunier}}{{Meunier}}{2018}]{Meunier_2018}
{Meunier} N.,  2018, \mn@doi [\aap] {10.1051/0004-6361/201730817}, \href
  {https://ui.adsabs.harvard.edu/abs/2018A&A...615A..87M} {615, A87}

\bibitem[\protect\citeauthoryear{{Meunier} \& {Lagrange}}{{Meunier} \&
  {Lagrange}}{2019}]{Meunier_2019}
{Meunier} N.,  {Lagrange} A.~M.,  2019, \mn@doi [\aap]
  {10.1051/0004-6361/201935099}, \href
  {https://ui.adsabs.harvard.edu/abs/2019A&A...625L...6M} {625, L6}

\bibitem[\protect\citeauthoryear{{Meunier} \& {Lagrange}}{{Meunier} \&
  {Lagrange}}{2020}]{Meunier_2020}
{Meunier} N.,  {Lagrange} A.~M.,  2020, \mn@doi [\aap]
  {10.1051/0004-6361/201937354}, \href
  {https://ui.adsabs.harvard.edu/abs/2020A&A...638A..54M} {638, A54}

\bibitem[\protect\citeauthoryear{{Meunier}, {Desort}  \& {Lagrange}}{{Meunier}
  et~al.}{2010a}]{Meunier_2010a}
{Meunier} N.,  {Desort} M.,   {Lagrange} A.~M.,  2010a, \mn@doi [\aap]
  {10.1051/0004-6361/200913551}, \href
  {https://ui.adsabs.harvard.edu/abs/2010A&A...512A..39M} {512, A39}

\bibitem[\protect\citeauthoryear{{Meunier}, {Lagrange}  \& {Desort}}{{Meunier}
  et~al.}{2010b}]{Meunier_2010b}
{Meunier} N.,  {Lagrange} A.~M.,   {Desort} M.,  2010b, \mn@doi [\aap]
  {10.1051/0004-6361/201014199}, \href
  {https://ui.adsabs.harvard.edu/abs/2010A&A...519A..66M} {519, A66}

\bibitem[\protect\citeauthoryear{{Meunier}, {Lagrange}, {Borgniet}  \&
  {Rieutord}}{{Meunier} et~al.}{2015}]{Meunier_2015}
{Meunier} N.,  {Lagrange} A.~M.,  {Borgniet} S.,   {Rieutord} M.,  2015,
  \mn@doi [\aap] {10.1051/0004-6361/201525721}, \href
  {https://ui.adsabs.harvard.edu/abs/2015A&A...583A.118M} {583, A118}

\bibitem[\protect\citeauthoryear{{Milbourne} et~al.,}{{Milbourne}
  et~al.}{2019}]{Milbourne_2019}
{Milbourne} T.~W.,  et~al., 2019, \mn@doi [\apj] {10.3847/1538-4357/ab064a},
  \href {https://ui.adsabs.harvard.edu/abs/2019ApJ...874..107M} {874, 107}

\bibitem[\protect\citeauthoryear{{Milbourne} et~al.,}{{Milbourne}
  et~al.}{2021}]{Milbourne_2021}
{Milbourne} T.~W.,  et~al., 2021, \mn@doi [\apj] {10.3847/1538-4357/ac1266},
  \href {https://ui.adsabs.harvard.edu/abs/2021ApJ...920...21M} {920, 21}

\bibitem[\protect\citeauthoryear{{Mortier}, {Faria}, {Correia}, {Santerne}  \&
  {Santos}}{{Mortier} et~al.}{2015}]{Mortier_2015}
{Mortier} A.,  {Faria} J.~P.,  {Correia} C.~M.,  {Santerne} A.,   {Santos}
  N.~C.,  2015, \mn@doi [\aap] {10.1051/0004-6361/201424908}, \href
  {https://ui.adsabs.harvard.edu/abs/2015A&A...573A.101M} {573, A101}

\bibitem[\protect\citeauthoryear{{Noyes}, {Hartmann}, {Baliunas}, {Duncan}  \&
  {Vaughan}}{{Noyes} et~al.}{1984}]{Noyes_1984}
{Noyes} R.~W.,  {Hartmann} L.~W.,  {Baliunas} S.~L.,  {Duncan} D.~K.,
  {Vaughan} A.~H.,  1984, \mn@doi [\apj] {10.1086/161945}, \href
  {https://ui.adsabs.harvard.edu/abs/1984ApJ...279..763N} {279, 763}

\bibitem[\protect\citeauthoryear{{Parker}}{{Parker}}{1978}]{Parker_1978}
{Parker} E.~N.,  1978, \mn@doi [\apj] {10.1086/156035}, \href
  {https://ui.adsabs.harvard.edu/abs/1978ApJ...221..368P} {221, 368}

\bibitem[\protect\citeauthoryear{{Pepe}, {Mayor}, {Galland}, {Naef}, {Queloz},
  {Santos}, {Udry}  \& {Burnet}}{{Pepe} et~al.}{2002}]{Pepe_2002}
{Pepe} F.,  {Mayor} M.,  {Galland} F.,  {Naef} D.,  {Queloz} D.,  {Santos}
  N.~C.,  {Udry} S.,   {Burnet} M.,  2002, \mn@doi [\aap]
  {10.1051/0004-6361:20020433}, \href
  {https://ui.adsabs.harvard.edu/abs/2002A&A...388..632P} {388, 632}

\bibitem[\protect\citeauthoryear{{Pepe} et~al.,}{{Pepe}
  et~al.}{2021}]{Pepe_2021}
{Pepe} F.,  et~al., 2021, \mn@doi [\aap] {10.1051/0004-6361/202038306}, \href
  {https://ui.adsabs.harvard.edu/abs/2021A&A...645A..96P} {645, A96}

\bibitem[\protect\citeauthoryear{{Pesnell}, {Thompson}  \&
  {Chamberlin}}{{Pesnell} et~al.}{2012}]{Pesnell_2012}
{Pesnell} W.~D.,  {Thompson} B.~J.,   {Chamberlin} P.~C.,  2012, \mn@doi
  [\solphys] {10.1007/s11207-011-9841-3}, \href
  {https://ui.adsabs.harvard.edu/abs/2012SoPh..275....3P} {275, 3}

\bibitem[\protect\citeauthoryear{{Phillips} et~al.,}{{Phillips}
  et~al.}{2016}]{Phillips_2016}
{Phillips} D.~F.,  et~al., 2016, in {Navarro} R.,  {Burge} J.~H.,  eds,
  Society of Photo-Optical Instrumentation Engineers (SPIE) Conference Series
  Vol. 9912, Advances in Optical and Mechanical Technologies for Telescopes and
  Instrumentation II. p. 99126Z, \mn@doi{10.1117/12.2232452}

\bibitem[\protect\citeauthoryear{{Pr{\v{s}}a} et~al.,}{{Pr{\v{s}}a}
  et~al.}{2016}]{IAU_2016}
{Pr{\v{s}}a} A.,  et~al., 2016, \mn@doi [\aj] {10.3847/0004-6256/152/2/41},
  \href {https://ui.adsabs.harvard.edu/abs/2016AJ....152...41P} {152, 41}

\bibitem[\protect\citeauthoryear{{Queloz} et~al.,}{{Queloz}
  et~al.}{2001}]{Queloz_2001}
{Queloz} D.,  et~al., 2001, \mn@doi [\aap] {10.1051/0004-6361:20011308}, \href
  {https://ui.adsabs.harvard.edu/abs/2001A&A...379..279Q} {379, 279}

\bibitem[\protect\citeauthoryear{{Quirrenbach} et~al.,}{{Quirrenbach}
  et~al.}{2016}]{Quirrenbach_2016}
{Quirrenbach} A.,  et~al., 2016, in {Evans} C.~J.,  {Simard} L.,   {Takami} H.,
   eds,  Society of Photo-Optical Instrumentation Engineers (SPIE) Conference
  Series Vol. 9908, Ground-based and Airborne Instrumentation for Astronomy VI.
  p. 990812, \mn@doi{10.1117/12.2231880}

\bibitem[\protect\citeauthoryear{{Radick}, {Lockwood}, {Henry}, {Hall}  \&
  {Pevtsov}}{{Radick} et~al.}{2018}]{Radick_2018}
{Radick} R.~R.,  {Lockwood} G.~W.,  {Henry} G.~W.,  {Hall} J.~C.,   {Pevtsov}
  A.~A.,  2018, \mn@doi [\apj] {10.3847/1538-4357/aaaae3}, \href
  {https://ui.adsabs.harvard.edu/abs/2018ApJ...855...75R} {855, 75}

\bibitem[\protect\citeauthoryear{{Reiners}}{{Reiners}}{2012}]{Reiners_2012}
{Reiners} A.,  2012, \mn@doi [Living Reviews in Solar Physics]
  {10.12942/lrsp-2012-1}, \href
  {https://ui.adsabs.harvard.edu/abs/2012LRSP....9....1R} {9, 1}

\bibitem[\protect\citeauthoryear{{Reiners} \& {Basri}}{{Reiners} \&
  {Basri}}{2006}]{Reiners_2006}
{Reiners} A.,  {Basri} G.,  2006, \mn@doi [\apj] {10.1086/503324}, \href
  {https://ui.adsabs.harvard.edu/abs/2006ApJ...644..497R} {644, 497}

\bibitem[\protect\citeauthoryear{{Rieutord}, {Roudier}, {Rincon}, {Malherbe},
  {Meunier}, {Berger}  \& {Frank}}{{Rieutord} et~al.}{2010}]{Rieutord_2010}
{Rieutord} M.,  {Roudier} T.,  {Rincon} F.,  {Malherbe} J.~M.,  {Meunier} N.,
  {Berger} T.,   {Frank} Z.,  2010, \mn@doi [\aap]
  {10.1051/0004-6361/200913303}, \href
  {https://ui.adsabs.harvard.edu/abs/2010A&A...512A...4R} {512, A4}

\bibitem[\protect\citeauthoryear{{Rincon} \& {Rieutord}}{{Rincon} \&
  {Rieutord}}{2018}]{Rincon_2018}
{Rincon} F.,  {Rieutord} M.,  2018, \mn@doi [Living Reviews in Solar Physics]
  {10.1007/s41116-018-0013-5}, \href
  {https://ui.adsabs.harvard.edu/abs/2018LRSP...15....6R} {15, 6}

\bibitem[\protect\citeauthoryear{{Robinson}}{{Robinson}}{1980}]{Robinson_1980}
{Robinson} R.~D. J.,  1980, \mn@doi [\apj] {10.1086/158184}, \href
  {https://ui.adsabs.harvard.edu/abs/1980ApJ...239..961R} {239, 961}

\bibitem[\protect\citeauthoryear{{Rueedi}, {Solanki}, {Livingston}  \&
  {Stenflo}}{{Rueedi} et~al.}{1992}]{Rueedi_1992}
{Rueedi} I.,  {Solanki} S.~K.,  {Livingston} W.,   {Stenflo} J.~O.,  1992,
  \aap, \href {https://ui.adsabs.harvard.edu/abs/1992A&A...263..323R} {263,
  323}

\bibitem[\protect\citeauthoryear{{Ryabchikova}, {Piskunov}, {Kurucz},
  {Stempels}, {Heiter}, {Pakhomov}  \& {Barklem}}{{Ryabchikova}
  et~al.}{2015}]{Ryabchikova_2015}
{Ryabchikova} T.,  {Piskunov} N.,  {Kurucz} R.~L.,  {Stempels} H.~C.,  {Heiter}
  U.,  {Pakhomov} Y.,   {Barklem} P.~S.,  2015, \mn@doi [\physscr]
  {10.1088/0031-8949/90/5/054005}, \href
  {https://ui.adsabs.harvard.edu/abs/2015PhyS...90e4005R} {90, 054005}

\bibitem[\protect\citeauthoryear{{Saar}}{{Saar}}{1988}]{Saar_1988}
{Saar} S.~H.,  1988, \mn@doi [\apj] {10.1086/165907}, \href
  {https://ui.adsabs.harvard.edu/abs/1988ApJ...324..441S} {324, 441}

\bibitem[\protect\citeauthoryear{{Saar}}{{Saar}}{2003}]{Saar_2003}
{Saar} S.~H.,  2003, in {Deming} D.,  {Seager} S.,  eds,  Astronomical Society
  of the Pacific Conference Series Vol. 294, Scientific Frontiers in Research
  on Extrasolar Planets. pp 65--70

\bibitem[\protect\citeauthoryear{{Saar}}{{Saar}}{2009}]{Saar_2009}
{Saar} S.~H.,  2009, in {Stempels} E.,  ed.,  American Institute of Physics
  Conference Series Vol. 1094, 15th Cambridge Workshop on Cool Stars, Stellar
  Systems, and the Sun. pp 152--161, \mn@doi{10.1063/1.3099086}

\bibitem[\protect\citeauthoryear{{Saar} \& {Donahue}}{{Saar} \&
  {Donahue}}{1997}]{Saar_1997}
{Saar} S.~H.,  {Donahue} R.~A.,  1997, \mn@doi [\apj] {10.1086/304392}, \href
  {https://ui.adsabs.harvard.edu/abs/1997ApJ...485..319S} {485, 319}

\bibitem[\protect\citeauthoryear{{Saar} \& {Linsky}}{{Saar} \&
  {Linsky}}{1985}]{Saar_1985}
{Saar} S.~H.,  {Linsky} J.~L.,  1985, \mn@doi [\apjl] {10.1086/184578}, \href
  {https://ui.adsabs.harvard.edu/abs/1985ApJ...299L..47S} {299, L47}

\bibitem[\protect\citeauthoryear{{Saar}, {Piskunov}  \& {Tuominen}}{{Saar}
  et~al.}{1992}]{Saar_1992}
{Saar} S.~H.,  {Piskunov} N.~E.,   {Tuominen} I.,  1992, in {Giampapa} M.~S.,
  {Bookbinder} J.~A.,  eds,  Astronomical Society of the Pacific Conference
  Series Vol. 26, Cool Stars, Stellar Systems, and the Sun. pp 255--258

\bibitem[\protect\citeauthoryear{{Scherrer} et~al.,}{{Scherrer}
  et~al.}{1995}]{Scherrer_1995}
{Scherrer} P.~H.,  et~al., 1995, \mn@doi [\solphys] {10.1007/BF00733429}, \href
  {https://ui.adsabs.harvard.edu/abs/1995SoPh..162..129S} {162, 129}

\bibitem[\protect\citeauthoryear{{Schou} et~al.,}{{Schou}
  et~al.}{2012}]{Schou_2012}
{Schou} J.,  et~al., 2012, \mn@doi [\solphys] {10.1007/s11207-011-9842-2},
  \href {https://ui.adsabs.harvard.edu/abs/2012SoPh..275..229S} {275, 229}

\bibitem[\protect\citeauthoryear{{Schrijver}, {Cote}, {Zwaan}  \&
  {Saar}}{{Schrijver} et~al.}{1989}]{Schrijver_1989}
{Schrijver} C.,  {Cote} J.,  {Zwaan} C.,   {Saar} S.,  1989, \mn@doi [\apj]
  {10.1086/167168}, \href
  {https://ui.adsabs.harvard.edu/abs/1989ApJ...337..964S} {337, 964}

\bibitem[\protect\citeauthoryear{{Seares}}{{Seares}}{1913}]{Seares_1913}
{Seares} F.~H.,  1913, \mn@doi [\apj] {10.1086/142014}, \href
  {https://ui.adsabs.harvard.edu/abs/1913ApJ....38...99S} {38, 99}

\bibitem[\protect\citeauthoryear{{Semel}}{{Semel}}{1989}]{Semel_1989}
{Semel} M.,  1989, \aap, \href
  {https://ui.adsabs.harvard.edu/abs/1989A&A...225..456S} {225, 456}

\bibitem[\protect\citeauthoryear{{Sheminova}}{{Sheminova}}{2019}]{Sheminova_2019}
{Sheminova} V.~A.,  2019, \mn@doi [Kinematics and Physics of Celestial Bodies]
  {10.3103/S088459131903005X}, \href
  {https://ui.adsabs.harvard.edu/abs/2019KPCB...35..129S} {35, 129}

\bibitem[\protect\citeauthoryear{{Skilling} \& {Bryan}}{{Skilling} \&
  {Bryan}}{1984}]{Skilling_1984}
{Skilling} J.,  {Bryan} R.~K.,  1984, \mn@doi [\mnras]
  {10.1093/mnras/211.1.111}, \href
  {https://ui.adsabs.harvard.edu/abs/1984MNRAS.211..111S} {211, 111}

\bibitem[\protect\citeauthoryear{{Skumanich} \& {L{\'o}pez Ariste}}{{Skumanich}
  \& {L{\'o}pez Ariste}}{2002}]{Skumanich_2002}
{Skumanich} A.,  {L{\'o}pez Ariste} A.,  2002, \mn@doi [\apj] {10.1086/339503},
  \href {https://ui.adsabs.harvard.edu/abs/2002ApJ...570..379S} {570, 379}

\bibitem[\protect\citeauthoryear{{Solanki}}{{Solanki}}{2003}]{Solanki_2003}
{Solanki} S.~K.,  2003, \mn@doi [\aapr] {10.1007/s00159-003-0018-4}, \href
  {https://ui.adsabs.harvard.edu/abs/2003A&ARv..11..153S} {11, 153}

\bibitem[\protect\citeauthoryear{{Spruit}}{{Spruit}}{1979}]{Spruit_1979}
{Spruit} H.~C.,  1979, \mn@doi [\solphys] {10.1007/BF00150420}, \href
  {https://ui.adsabs.harvard.edu/abs/1979SoPh...61..363S} {61, 363}

\bibitem[\protect\citeauthoryear{{Stenflo}}{{Stenflo}}{2013}]{Stenflo_2013}
{Stenflo} J.~O.,  2013, \mn@doi [\aapr] {10.1007/s00159-013-0066-3}, \href
  {https://ui.adsabs.harvard.edu/abs/2013A&ARv..21...66S} {21, 66}

\bibitem[\protect\citeauthoryear{{Stenflo} \& {Lindegren}}{{Stenflo} \&
  {Lindegren}}{1977}]{Stenflo_1977}
{Stenflo} J.~O.,  {Lindegren} L.,  1977, \aap, \href
  {https://ui.adsabs.harvard.edu/abs/1977A&A....59..367S} {59, 367}

\bibitem[\protect\citeauthoryear{{Stenflo}, {Solanki}, {Harvey}  \&
  {Brault}}{{Stenflo} et~al.}{1984}]{Stenflo_1984}
{Stenflo} J.~O.,  {Solanki} S.,  {Harvey} J.~W.,   {Brault} J.~W.,  1984, \aap,
  \href {https://ui.adsabs.harvard.edu/abs/1984A&A...131..333S} {131, 333}

\bibitem[\protect\citeauthoryear{{Takeda} \& {UeNo}}{{Takeda} \&
  {UeNo}}{2017}]{Takeda_2017}
{Takeda} Y.,  {UeNo} S.,  2017, \mn@doi [\pasj] {10.1093/pasj/psx022}, \href
  {https://ui.adsabs.harvard.edu/abs/2017PASJ...69...46T} {69, 46}

\bibitem[\protect\citeauthoryear{{Thompson}, {Watson}, {de Mooij}  \&
  {Jess}}{{Thompson} et~al.}{2017}]{Thompson_2017}
{Thompson} A.~P.~G.,  {Watson} C.~A.,  {de Mooij} E.~J.~W.,   {Jess} D.~B.,
  2017, \mn@doi [\mnras] {10.1093/mnrasl/slx018}, \href
  {https://ui.adsabs.harvard.edu/abs/2017MNRAS.468L..16T} {468, L16}

\bibitem[\protect\citeauthoryear{{Thompson} et~al.,}{{Thompson}
  et~al.}{2020}]{Thompson_2020}
{Thompson} A.~P.~G.,  et~al., 2020, \mn@doi [\mnras] {10.1093/mnras/staa1010},
  \href {https://ui.adsabs.harvard.edu/abs/2020MNRAS.494.4279T} {494, 4279}

\bibitem[\protect\citeauthoryear{{Title} \& {Tarbell}}{{Title} \&
  {Tarbell}}{1975}]{Title_1975}
{Title} A.~M.,  {Tarbell} T.~D.,  1975, \mn@doi [\solphys]
  {10.1007/BF00154064}, \href
  {https://ui.adsabs.harvard.edu/abs/1975SoPh...41..255T} {41, 255}

\bibitem[\protect\citeauthoryear{{Ulmer-Moll}, {Figueira}, {Neal}, {Santos}  \&
  {Bonnefoy}}{{Ulmer-Moll} et~al.}{2019}]{UlmerMoll_2019}
{Ulmer-Moll} S.,  {Figueira} P.,  {Neal} J.~J.,  {Santos} N.~C.,   {Bonnefoy}
  M.,  2019, \mn@doi [\aap] {10.1051/0004-6361/201833282}, \href
  {https://ui.adsabs.harvard.edu/abs/2019A&A...621A..79U} {621, A79}

\bibitem[\protect\citeauthoryear{{Valenti}, {Marcy}  \& {Basri}}{{Valenti}
  et~al.}{1995}]{Valenti_1995}
{Valenti} J.~A.,  {Marcy} G.~W.,   {Basri} G.,  1995, \mn@doi [\apj]
  {10.1086/175231}, \href
  {https://ui.adsabs.harvard.edu/abs/1995ApJ...439..939V} {439, 939}

\bibitem[\protect\citeauthoryear{{Wiegelmann}, {Thalmann}  \&
  {Solanki}}{{Wiegelmann} et~al.}{2014}]{Wiegelmann_2014}
{Wiegelmann} T.,  {Thalmann} J.~K.,   {Solanki} S.~K.,  2014, \mn@doi [\aapr]
  {10.1007/s00159-014-0078-7}, \href
  {https://ui.adsabs.harvard.edu/abs/2014A&ARv..22...78W} {22, 78}

\bibitem[\protect\citeauthoryear{{Yeo}, {Solanki}  \& {Krivova}}{{Yeo}
  et~al.}{2013}]{Yeo_2013}
{Yeo} K.~L.,  {Solanki} S.~K.,   {Krivova} N.~A.,  2013, \mn@doi [\aap]
  {10.1051/0004-6361/201220682}, \href
  {https://ui.adsabs.harvard.edu/abs/2013A&A...550A..95Y} {550, A95}

\bibitem[\protect\citeauthoryear{{Yu}, {Huber}, {Bedding}  \& {Stello}}{{Yu}
  et~al.}{2018}]{Yu_2018}
{Yu} J.,  {Huber} D.,  {Bedding} T.~R.,   {Stello} D.,  2018, \mn@doi [\mnras]
  {10.1093/mnrasl/sly123}, \href
  {https://ui.adsabs.harvard.edu/abs/2018MNRAS.480L..48Y} {480, L48}

\bibitem[\protect\citeauthoryear{{Zeeman}}{{Zeeman}}{1897}]{Zeeman_1897}
{Zeeman} P.,  1897, \nat

\bibitem[\protect\citeauthoryear{{Zhao} et~al.,}{{Zhao}
  et~al.}{2022}]{Zhao_2022}
{Zhao} L.~L.,  et~al., 2022, \mn@doi [\aj] {10.3847/1538-3881/ac5176}, \href
  {https://ui.adsabs.harvard.edu/abs/2022AJ....163..171Z} {163, 171}

\bibitem[\protect\citeauthoryear{{Zirin} \& {Popp}}{{Zirin} \&
  {Popp}}{1989}]{Zirin_1989}
{Zirin} H.,  {Popp} B.,  1989, \mn@doi [\apj] {10.1086/167418}, \href
  {https://ui.adsabs.harvard.edu/abs/1989ApJ...340..571Z} {340, 571}

\makeatother
\end{thebibliography}




\appendix

\section{Expansion in Hermite-Gaussian polynomials} \label{app:expansion}

\begin{equation} \label{eq:hgrv0}
    g(\lambda;\xi) = e^{-\frac{(\lambda-\mu)^2}{2\sigma_{\lambda}^2}}-e^{-\frac{(\xi \lambda-\mu)^2}{2\sigma_{\lambda}^2}}
\end{equation}

can be expanded as

\begin{equation}
        \sum_{n=0}^{\infty}c_n(\xi)\psi_n(\lambda;\mu,\sigma_{\lambda})
\end{equation}
with $\epsilon = \xi-1$.

For n = 0:
\begin{equation} \label{Aeq:coeff0}
    c_{0}(\epsilon)=\sqrt{\sigma_{\lambda} \sqrt{\pi}}-\frac{1}{\sqrt{\sigma_{\lambda} \sqrt{\pi}}} I_{0}\left(\frac{1+\epsilon+\frac{\epsilon^{2}}{2}}{\sigma_{\lambda}^{2}},-\frac{2 \mu+\epsilon \mu}{\sigma_{\lambda}^{2}},\left(\frac{\mu}{\sigma_{\lambda}}\right)^{2}\right)
\end{equation}

for n $\geq$ 1:


\begin{multline}\label{Aeq:coeffn}
   c_{n}(\epsilon)=-\sqrt{\frac{\sigma_{\lambda} n ! 2^{n}}{\sqrt{\pi}}} \sum_{m=0}^{\left\lfloor\frac{n}{2}\right\rfloor} \frac{(-1)^{m}}{4^{m} m !(n-2 m) !} \\ I_{n-2 m}\left(1+\epsilon+\frac{\epsilon^{2}}{2}, \frac{\epsilon \mu}{\sigma_{\lambda}}(1+\epsilon), \frac{1}{2}\left(\frac{\epsilon \mu}{\sigma_{\lambda}}\right)^{2}\right)
\end{multline} 

with

\begin{equation} \label{Aeq:hg}
    \psi_{n}(\lambda ; \mu, \sigma_{\lambda})=\frac{1}{\sqrt{\sigma_{\lambda} 2^{n} n ! \sqrt{\pi}}} H_{n}\left(\frac{\lambda-\mu}{\sigma_{\lambda}}\right) e^{-\frac{(\lambda-\mu)^{2}}{2 \sigma_{\lambda}^{2}}}
\end{equation}
\vspace{1cm}

$H_n$ is the nth degree physicist's Hermite polynomial:\\
$H_0(s) = 1$ \\
$H_1(s) = 2s$ \\
$H_2(s) = 4s^2-2$ \\
$H_3(s) = 8s^3-12s$ \\
\vspace{0.3cm}

$I_n$ is defined as:
\begin{equation}
I_{0}(a, b, c)=\sqrt{\frac{\pi}{a}} e^{\left(\frac{b^{2}}{4 a}-c\right)}
\end{equation}
\begin{equation}
I_{1}(a, b, c)=-\frac{\sqrt{\pi} b}{2 a^{3 / 2}} e^{\left(\frac{b^{2}}{4 a}-c\right)}
\end{equation}

\begin{equation}
I_{n}(a, b, c)=-\frac{b}{2 a} I_{n-1}(a, b, c)+\frac{n-1}{2 a} I_{n-2}(a, b, c)
\end{equation}

for all $n\geq2$.

\section{Residual profile caused by simple broadening} \label{broadening_residual_profile}
Taylor expand Eq. \ref{eq:sigma_broaden} for r near 0:
\begin{equation}
    d e^{-\frac{(v-v_0)^2}{2\sigma_{v}^2}}
    -d e^{-\frac{(v-v_0)^2}{2\sigma_{v}^2}}
    +d e^{-\frac{(v-v_0)^2}{2\sigma_{v}^2}} r
    -d \frac{(v-v_0)^2}{\sigma_{v}^2} e^{-\frac{(v-v_0)^2}{2\sigma_{v}^2}} r
\end{equation}
which is equal to
\begin{equation} \label{eq:26_approx}
    d \left(r
    - r\frac{(v-v_0)^2}{\sigma_{v}^2} \right) e^{-\frac{(v-v_0)^2}{2\sigma_{v}^2}}.
\end{equation}
Eq. 14 on the other hand was:
\begin{equation} 
    I_{\text{diff}}(v) = 2d\alpha\epsilon^2\left(\frac{c^2}{2\sigma_{v}^2} +\frac{3}{8} - \left(1+\frac{c^2}{2\sigma_{v}^2}\right) \left(\frac{v-v_0}{\sigma_{v}}\right)^2\right)e^{-\frac{(v-v_0)^{2}}{2 \sigma_{v}^{2}}}.
\end{equation} 
which is approximately equal to (keeping only the dominant terms, $\frac{c}{\sigma_v}>>1$):
\begin{equation} \label{eq:idiff_v_approx}
    I_{\text{diff}}(v) = d\alpha\epsilon^2\left(\frac{c^2}{\sigma_{v}^2}  - \frac{c^2}{\sigma_{v}^2} \left(\frac{v-v_0}{\sigma_{v}}\right)^2\right)e^{-\frac{(v-v_0)^{2}}{2 \sigma_{v}^{2}}}.
\end{equation}

Eq. \ref{eq:26_approx} and \ref{eq:idiff_v_approx} are equal for
\begin{equation} \label{Eq.r_factor}
    r = \alpha\epsilon^2 \frac{c^2}{\sigma_{v}^2}.
\end{equation}
Therefore, the residual profile from the broadened Gaussians matches the residual profile computed using the $\pi$ and $\sigma$ components for $r$ set to the value in Eq. \ref{Eq.r_factor}. It follows that for small variations of the magnetic field, it does not matter whether we model the Zeeman effect as simple broadening or splitting, as the residual shapes are the same and \dabb can be extracted from either residual profile.

\section{Correlations with high activity data chunks} \label{app:correlations_150cm}
\begin{figure}
    \centering
    \includegraphics[width=\columnwidth]{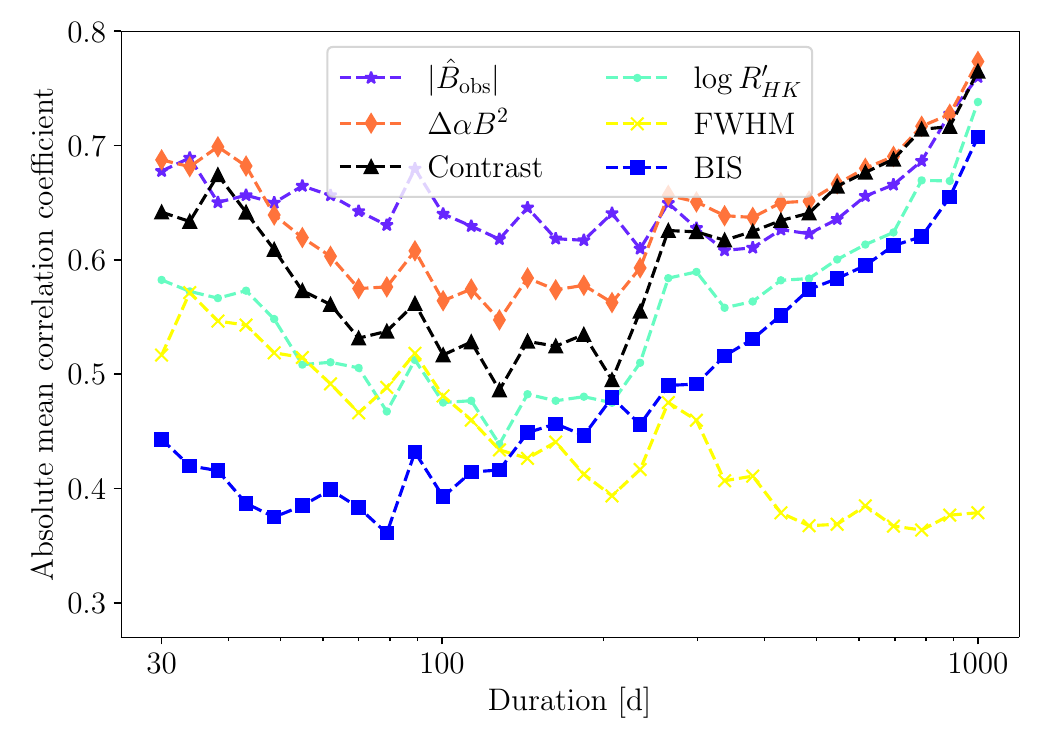}
    \caption{Absolute mean Pearson correlation coefficient of the six activity indices with the heliocentric CCF RVs for data chunks covering a fixed time span.}
    \label{fig:rvcorrs150}
\end{figure}

\begin{figure}
    \centering
    \includegraphics[width=\columnwidth]{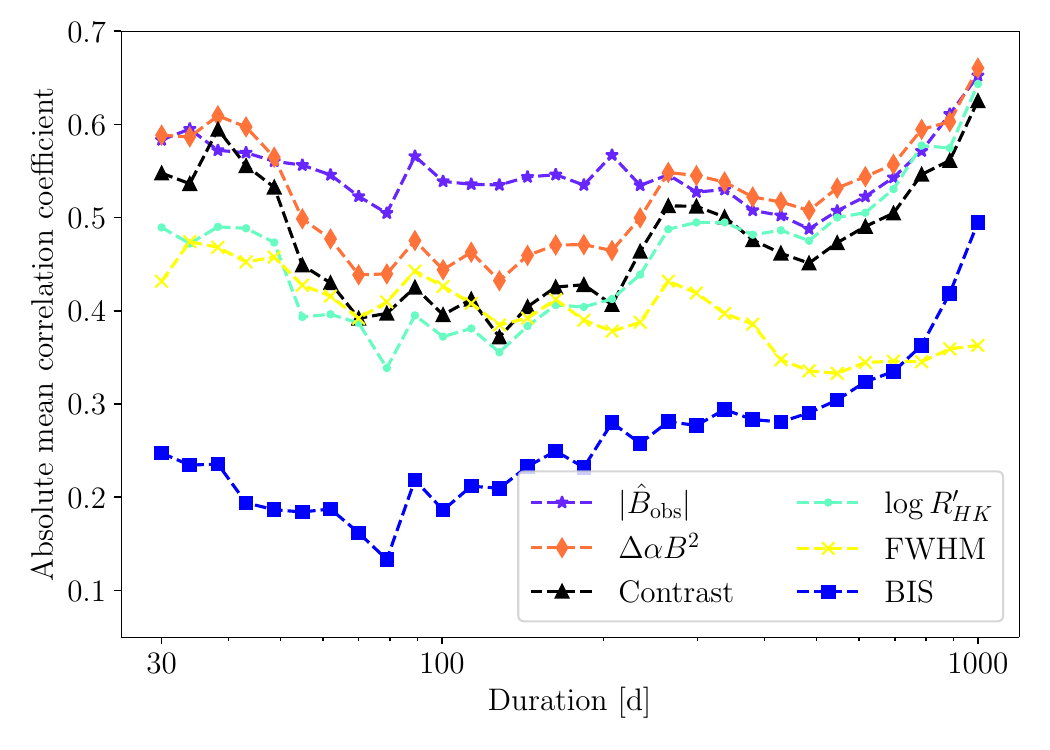}
    \caption{Absolute mean Pearson correlation coefficient of the six activity indices with the heliocentric MM-LSD RVs for data chunks covering a fixed time span.}
    \label{fig:rvcorrs_lsd150}
\end{figure}

\begin{figure}
    \centering
    \includegraphics[width=\columnwidth]{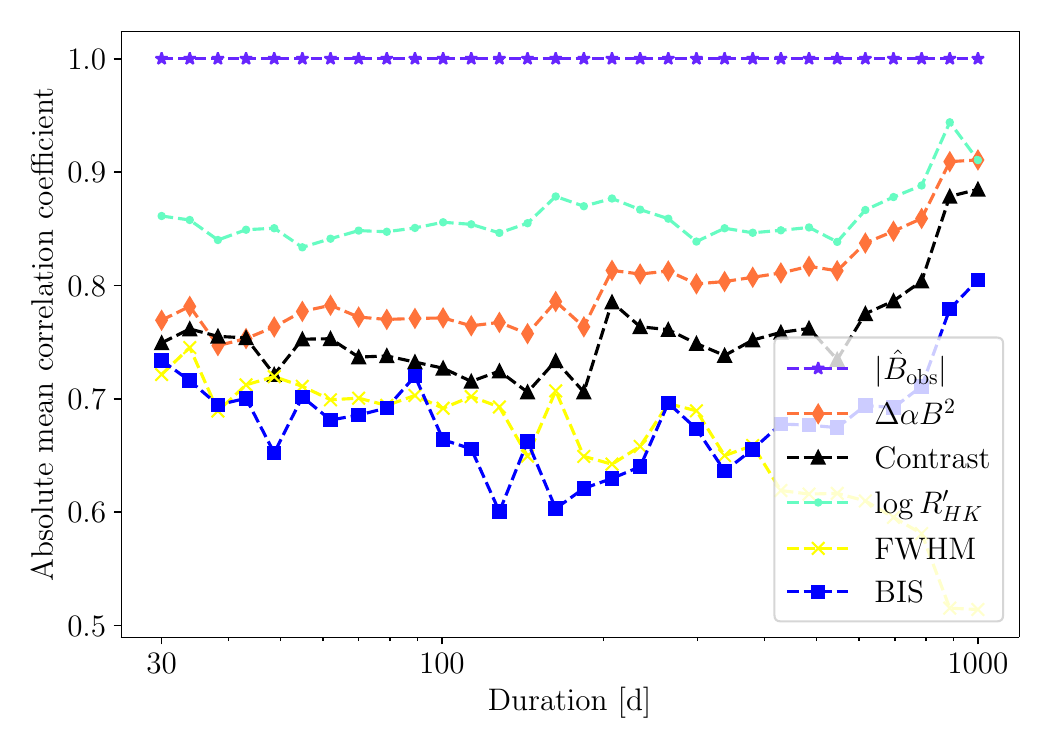}
    \caption{Absolute mean Pearson correlation coefficient of the five activity indices derived from HARPS-N spectra with \bobs for data chunks covering a fixed time span.}
    \label{fig:bcorrs150}
\end{figure}

The Figs. \ref{fig:rvcorrs150}, \ref{fig:rvcorrs_lsd150}, and \ref{fig:bcorrs150} show the same correlation analyses as those shown in Figs. \ref{fig:rvcorrs}, \ref{fig:rvcorrs_lsd}, and \ref{fig:bcorrs}, but with the threshold for included data chunks set to 1.5 \ms.

\bsp	
\label{lastpage}
\end{document}